\documentclass[12pt]{article}
\usepackage{theorem}
\usepackage{amsmath,amssymb}
\usepackage{graphicx,makeidx}
\usepackage{caption,subcaption}
\usepackage[mathscr]{eucal}
\usepackage{setspace}

\usepackage{titlesec}
\titleformat{\subsection}[runin]
 {\normalfont\normalsize\bfseries}{\thesubsection}{1em}{\!\!}
\titleformat{\subsubsection}[runin]
 {\normalfont\normalsize\bfseries}{\thesubsubsection}{1em}{\!\!}
{\theorembodyfont{\upshape} }
\newcommand{\boma}[1]{{\mbox{\boldmath $#1$} }}

\begin{document}
\def\IE{\mathfrak{I}}
\def\rg{\mathfrak r}
\def\raL{\ra}
\def\Cc{F}
\def\Dc{G}
\def\E0{E_0}
\def\Dik{\Dir^{(k)}}
\def\vi{z}
\def\zi{v}
\def\vt{v}
\def\En{\mathfrak{E}}
\def\TAM{A_{(>)}}
\def\TAm{A_{(<)}}
\def\InB{\mathfrak{B}^\s}
\def\eu{\varepsilon^{\s}}
\def\eren{\varepsilon^{ren}}
\def\TT{\mathscr{T}}
\def\TTs{\TT^{\s}}
\def\an{\lambda}
\def\ET{{}_T\mathcal{E}}
\def\Ep{{}_p\mathcal{E}}
\def\b0{\textbf{0}}
\def\Ta{\mathsf{T}}
\def\Bp{{B_{\sp,n}}}
\def\Bs{{B_{\si,n}}}
\def\sp{{\bar{\si}}}
\def\Sp{{\bar{S}}}
\def\ba{{\bf a}}
\def\bap{\mathfrak{a}}
\def\sip{\mathfrak{p}}
\def\bj{\sip(j)}
\def\Op{\mathfrak{O}}
\def\fo{\mathfrak{F}}
\def\HE{H}
\def\uu{v}
\def\AM{A}
\def\Am{a}
\def\DD{D}
\def\blp{\bar{\bf l}}
\def\bkl{b_{\bk\bl}}
\def\bklp{\bar{b}_{\bk\blp}}
\def\bxp{\bar{\bx}}
\def\bkp{\bar{\bk}}
\def\CB{C}
\def\aa{\lambda}
\def\bxk{\bx_{\star}}
\def\xk{x_\star}
\def\Co{\lozenge}
\def\NCo{\blacksquare}
\def\Toro{{\bf T}}
\def\atanh{\mbox{arcth}\,}
\def\Iun{\mathfrak{J}}
\def\GaL{\mathfrak{g}}
\def\Mk{M_{\mm,k}}
\def\ri{\mathfrak{r}}
\def\oms{\omega_{*}}
\def\Fock{\mathfrak{F}}
\def\Ee{\mathfrak{E}}
\def\ee{\epsilon}
\def\Ome{\Om_\ee}
\def\bxe{\bx_\ee}
\def\FI{\boma{S}}
\def\dFI{\mathfrak{S}}
\def\ff{\mathcal{F}}
\def\hh{\mathcal{H}}
\def\Mel{\mathfrak{M}}
\def\Tr{\mbox{Tr}\,}
\def\xb{\bar{x}}
\def\rb{\bar{\rho}}
\def\rr{\bar{r}}
\def\uno{\mbox{\textbf{1}}}
\def\tez{\al}
\def\Res{\mbox{Res}}
\def\L2m0{L^2_0}
\def\Tm{\tilde{T}}
\def\II{\mathfrak{I}}
\def\Ns{\mathfrak{N}}
\def\Tti{\tilde{T}}
\def\DF{\mathscr{D}}
\def\F{{\mathcal F}}
\def\mm{\kappa}
\def\oo{\varpi}
\def\0{{\bf 0}}
\def\UU{{\mathcal U}}
\def\Hank{\mathfrak{H}}
\def\t{\mathfrak{t}}
\def\Dir{D}
\def\l{\left}
\def\r{\right}
\def\ha{\widehat{a}}
\def\ak{\ha_k}
\def\ah{\ha_h}
\def\had{\ha^{\dagger}}
\def\akd{\had_k}
\def\ahd{\had_h}
\def\GD{\mathfrak{G}}
\def\Fk{F_k}
\def\Fh{F_h}
\def\Fkc{\overline{\Fk}}
\def\Fhc{\overline{\Fh}}
\def\Fkd{\mathfrak{F}_{k_1}}
\def\fk{f_k}
\def\fh{f_h}
\def\fkc{\overline{\fk}}
\def\fhc{\overline{\fh}}
\def\bl{{\bf l}}
\def\bn{{\bf n}}
\def\bk{{\bf k}}
\def\bh{{\bf h}}
\def\bx{{\bf x}}
\def\by{{\bf y}}
\def\bz{{\bf z}}
\def\bq{{\bf q}}
\def\bp{{\bf p}}
\def\Nab{\square}
\def\Fi{\widehat{\phi}}
\def\s{u}
\def\Fis{\Fi^{\s}}
\def\Fiseps{\Fi^{\eps \s}}
\def\Ti{\widehat{T}}
\def\Tis{\Ti^{\s}}
\def\Tiseps{\Ti^{\eps \s}}
\def\Aa{\widehat{A}}
\def\Bb{\widehat{B}}
\def\al{\alpha}
\def\be{\beta}
\def\de{\delta}
\def\eps{\varepsilon}
\def\ga{\gamma}
\def\lam{\lambda}
\def\om{\omega}
\def\si{\sigma}
\def\te{\theta}
\def\Ga{\Gamma}
\def\Om{\Omega}
\def\Si{\Sigma}
\def\dd{\displaystyle}
\def\la{\langle}
\def\ra{\rangle}
\def\leqs{\leqslant}
\def\geqs{\geqslant}
\def\sc{\cdot}
\def\restriction{\upharpoonright}
\def\parn{\par\noindent}
\def\complessi{{\bf C}}
\def\reali{{\bf R}}
\def\razionali{{\bf Q}}
\def\interi{{\bf Z}}
\def\naturali{{\bf N}}
\def\AA{{\mathcal A}}
\def\BB{{\mathcal B}}
\def\FF{{\mathcal F}}
\def\EE{{\mathcal E}}
\def\GG{{\mathcal G} }
\def\HH{{\mathcal H}}
\def\JJ{{\mathcal J}}
\def\KK{{\mathcal K}}
\def\LL{{\mathcal L}}
\def\MM{{\mathcal M}}
\def\OO{{\mathcal O}}
\def\PP{{\mathcal P}}
\def\QQ{{\mathcal Q}}
\def\RR{{\mathcal R}}
\def\SS{{\mathcal S}}
\def\cir{{\scriptscriptstyle \circ}}
\def\parn{\par \noindent}
\def\salto{\vskip 0.2truecm \noindent}
\def\beq{\begin{equation}}
\def\feq{\end{equation}}
\def\barray{\begin{array}}
\def\farray{\end{array}}
\newcommand{\rref}[1]{(\ref{#1})}
\def\vain{\rightarrow}
\def\spazio{\vskip 0.5truecm \noindent}
\def\fine{\hfill $\square$ \vskip 0.2cm \noindent}
\def\ffine{\hfill $\lozenge$ \vskip 0.2cm \noindent}

\setcounter{secnumdepth}{5}

\makeatletter \@addtoreset{equation}{section}
\renewcommand{\theequation}{\thesection.\arabic{equation}}
\makeatother
\begin{titlepage}
\begin{center}
{\Large \textbf{Local zeta regularization}}
\vskip 0.2cm
{~}
\hskip -0.4cm
{\Large \textbf{and the scalar Casimir effect III.}}
\vskip 0.2cm
{~}
\hskip -0.4cm
{\Large\textbf{The case with a background harmonic potential}}
\end{center}
\vspace{0.5truecm}
\begin{center}
{\large
Davide Fermi$\,{}^a$, Livio Pizzocchero$\,{}^b$({\footnote{Corresponding author}})} \\
\vspace{0.5truecm}
${}^a$ Dipartimento di Matematica, Universit\`a di Milano\\
Via C. Saldini 50, I-20133 Milano, Italy\\
e--mail: davide.fermi@unimi.it \\
\vspace{0.2truecm}
${}^b$ Dipartimento di Matematica, Universit\`a di Milano\\
Via C. Saldini 50, I-20133 Milano, Italy\\
and Istituto Nazionale di Fisica Nucleare, Sezione di Milano, Italy \\
e--mail: livio.pizzocchero@unimi.it
\end{center}
\begin{abstract}
Applying the general framework for local zeta regularization
proposed in Part I of this series of papers, we renormalize
the vacuum expectation value of the stress-energy tensor (and of the total energy) for a scalar field
in presence of an external harmonic potential.
\end{abstract}
\vspace{0.2cm} \noindent
\hfill \parn
\par \vspace{0.3truecm} \noindent \textbf{AMS Subject classifications:} 81T55, 83C47.
\par \vspace{0.3truecm} \noindent \textbf{PACS}: 03.70.+k, 11.10.Gh, 41.20.Cv~.
\end{titlepage}
\begin{spacing}{0.95}
  \tableofcontents
\end{spacing}
\vfill \eject \noindent
\section{Introduction}
In Part I of this series of papers \cite{PI,PII,PIV} (partly inspired by
\cite{ptp}) we have considered the
general framework of local (and global) zeta regularization of a neutral
scalar field in a $d$-dimensional spatial domain $\Om$, in the environment
of $(d+1)$-dimensional Minkowski spacetime; the possible presence of an
external potential $V$ was taken into account as well. \parn
In the present Part III we consider a massless field on $\Om = \reali^d$,
and choose for $V$ a harmonic potential. The potential is assumed to
be isotropic, i.e., to be proportional to the squared radius $|\bx|^2$
($\bx \in \reali^d$); nevertheless, our approach could be extended
with little effort to anisotropic harmonic potentials and to the case
of a massive scalar field. \parn
The main result of this paper is the renormalization of the vacuum
expectation value (VEV) for the stress-energy tensor, obtained applying
the general framework of Part I to the present configuration; we also
discuss the total energy in this configuration. After producing general
integral representations for both the renormalized stress-energy VEV
and the total energy, we consider in detail the cases of spatial dimension
$d \in \{1,2,3\}$. \parn
The above mentioned integral representations are derived analytically
and are fully explicit; however, the integrals therein must be computed
numerically. We have performed these latter computations for $d \in \{1,2,3\}$,
using $\tt{Mathematica}$. \parn
Using the previously cited integral representation, it is also possible
to derive the asymptotics for the stress-energy tensor components when
the radius $|\bx|$ goes to zero or to infinity; we derive, as well,
remainder estimates for these asymptotic expansions as well. \parn
The idea to replace sharp boundaries with suitable background potentials
is well-known in the literature on the Casimir effect. Typically
(see, e.g., \cite{delp4,delp1,delp2,MaTru4,delp3}), delta-like potentials
are introduced in order to mimic boundary conditions in a ``physically
more realistic'' framework; the ultimate purpose is to obtain less singular
behaviours of the renormalized quantities, avoiding, e.g., boundary
divergences such as the ones pointed out in Parts I and II \cite{PI,PII}.
The case of a scalar field interacting with an external harmonic
potential has been formerly considered by Actor and Bender
\cite{ActHarm1,ActHarm2}, who have determined the renormalized VEV
of the total energy via global zeta regularization; to the best of
our knowledge, local aspects such as the stress-energy tensor have
never been considered previously for the present configuration. \parn
The paper is organized as follows. In Section \ref{Gloss} we present
a summary of basic results from Part I; as in Part II, the purpose
of the summary is to save the reader from the tedious task of recovering
from Part I the general identities applied here. In Section \ref{secHarm}
(and in the related Appendix \ref{AppII}) we discuss the general approach
to treat the case of a scalar field interacting with a classical, isotropic
harmonic potential in arbitrary spatial dimension $d$; this includes general
expressions for the renormalized stress-energy VEV, its asymptotics for
$|\bx|$ small or large and for the renormalized VEV of the total energy
(more precisely, for the bulk energy as defined in Part I).
The main ingredients in the derivation of the above results are: \parn
i) the general results of Part I relating the zeta regularization
to the heat kernel of fundamental operator $\AA = - \Delta + V(\bx)$; \parn
ii) the fact that, when $V(\bx)$ is harmonic, the related heat kernel
is the well-known Mehler kernel \cite{Mehl} (also see \cite{Berl,Calin,Dav,Gryg}). \parn
In Section \ref{Harmd123}, the analysis presented in the preceeding
sections is specified to the cases of spatial dimension $d \in \{1,2,3\}$. \parn
We point out that results on the renormalized bulk energy agree
with the ones of Actor and Bender \cite{ActHarm2} for $d \in \{1,2,3\}$.
Nevertheless, it should be mentioned that these authors compute
the analytic continuations required by the zeta approach using
ad hoc results on continuation the special functions involved
in this specific case; on the contrary, here we are applying
mechanically the general setting discussed in Part I,
in the spirit of the present series of papers.
\section{A summary of results from Part I}\label{Gloss}
\subsection{General setting.}
Throughout the paper we use natural units, so that
\beq c = 1 ~, \qquad \hbar = 1 ~. \feq
Our approach works in $(d+1)$-dimensional Minkowski spacetime, which is
identified with $\reali^{d+1}$ using a set of inertial coordinates
\beq x = (x^\mu)_{\mu=0,1,...,d} \equiv (x^0,\bx) \equiv (t,\bx) ~; \feq
the Minkowski metric is $(\eta_{\mu \nu}) = \mbox{diag} (-1,1,...\,,1)$\,.
We fix a spatial domain $\Om \subset \reali^d$ and a background static
potential $V\!:\!\Om\!\to\!\reali$. We consider a quantized neutral,
scalar field $\Fi : \reali \times \Om \to \LL_{s a}(\Fock)$ ($\Fock$
is the Fock space and $\LL_{s a}(\Fock)$ are the selfadjoint operators on it);
suitable boundary conditions are prescribed on $\partial \Om$. The field
equation reads
\beq 0 = (-\partial_{tt}+\Delta-V(\bx)) \Fi(\bx,t) \label{daquan} \feq
($\Delta := \sum_{i=1}^d \partial_{ii}$ is the $d$-dimensional Laplacian).
We put
\beq \AA := - \Delta + V ~, \label{defaa} \feq
keeping into account the boundary conditions on $\partial\Om$, and consider
the Hilbert space $L^2(\Om)$ with inner product $\la f|g\raL :=
\int_{\Om}d\bx\,\overline{f}(\bx)g(\bx)$. We assume $\AA$ to be selfadjoint
in $L^2(\Om)$ and strictly positive (i.e., with spectrum $\si(\AA) \subset
[\eps^2,+\infty)$ for some $\eps >0$). \parn
We often refer to a complete orthonormal set $(\Fk)_{k \in \KK}$ of
(proper or improper) eigenfunctions of $\AA$ with eigenvalues
$(\om^2_k)_{k \in \KK}$ ($\om_k\!\geqs\!\eps$ for all $k\!\in\!\KK$). Thus
\begin{equation}\begin{split}
& \Fk : \Om \to \complessi; \qquad \AA\Fk= \om^2_k \Fk ~; \\
& \hspace{-0.5cm} \la \Fk | \Fh \raL = \de(k, h) \quad
\mbox{for all $k,h \in \KK$} ~. \label{eigenf}
\end{split}\end{equation}
The labels $k\!\in\!\KK$ can include both discrete and continuous
parameters; $\int_\KK dk$ indicates summation over all labels and
$\de(h,k)$ is the Dirac delta function on $\KK$. \parn
We expand the field $\Fi$ in terms of destruction and creation
operators corresponding to the above eigenfunctions, and assume the
canonical commutation relations; $|0 \ra \in \Fock$ is the vacuum
state and VEV stands for ``vacuum expectation value''. \parn
The quantized stress-energy tensor reads ($\xi\!\in\! \reali$ is a parameter)
\beq \Ti_{\mu \nu} := \l(1 - 2\xi \r) \partial_\mu \Fi \circ \partial_\nu \Fi
- \l({1\over 2} - 2\xi \r)\eta_{\mu\nu}(\partial^\lam\Fi \, \partial_\lam \Fi + V\Fi^2)
- 2 \xi \, \Fi \circ \partial_{\mu \nu} \Fi \,; \label{tiquan} \feq
in the above we put $\Aa \circ \Bb := (1/2) (\Aa \Bb + \Bb \Aa)$
for all $\Aa, \Bb \in \LL_{s a}(\Fock)$, and all the bilinear terms
in the field are evaluated on the diagonal (e.g., $\partial_\mu \Fi
\circ \partial_\nu \Fi$ indicates the map $x \mapsto \partial_\mu \Fi(x)
\circ \partial_\nu \Fi(x)$). The VEV $\la 0|\Ti_{\mu\nu}|0\ra$ is
typically divergent.
\vspace{-0.4cm}
\subsection{Zeta regularization.}
The \textsl{zeta-regularized field operator} is
\beq \Fis := (\mm^{-2} \AA)^{-\s/4} \Fi ~, \label{Fis} \feq
where $\AA$ is the operator \rref{defaa}, $\s \in \complessi$
and $\mm > 0$ is a ``mass scale'' parameter; note that
$\Fis|_{\s = 0} = \Fi$, at least formally. The \textsl{zeta
regularized stress-energy tensor} is
\beq \Tis_{\mu \nu} := (1\!-\!2\xi)\partial_\mu \Fis\!\circ\partial_\nu \Fis\!
- \!\l({1\over 2}\!-\!2\xi\!\r)\!\eta_{\mu\nu}\!
\l(\!\partial^\lam\Fis\partial_\lam \Fis\!+\!V(\Fis)^2\r)
- 2 \xi \, \Fis\!\circ \partial_{\mu \nu} \Fis \,. \label{tiquans} \feq
The VEV $\la 0|\Tis_{\mu\nu}|0\ra$ is well defined for $\Re\s$ large enough
(see the forthcoming subsection \ref{DirT}); moreover, in the region of definition
it is an analytic function of $\s$. The same can be said of many related observables
(including global objects, such as the total energy VEV). \parn
For any one of these observables, let us denote with $\F(\s)$ its
zeta-regularized version and assume this to be analytic for $\s$
in a suitable domain $\UU_0 \subset \complessi$.
The zeta approach to renormalization can be formulated in two versions. \parn
i) \textsl{Restricted version}. Assume the map $\UU_0\!\to\!\complessi$,
$\s\!\mapsto\!\F(\s)$ to admit an analytic continuation (indicated with the
same notation) to an open subset $\UU\!\subset\!\complessi$ with $\UU\!\ni\!0$\,;
then we define the renormalized observables as
\beq \F_{ren} := \F(0) ~. \label{ren} \feq
ii) \textsl{Extended version}. Assume that there exists an open subset
$\UU\!\subset\!\complessi$ with $\UU_0\!\subset\!\UU$, such that
$0\!\in\!\UU$ and the map $\s\!\in\!\UU_0 \mapsto \F(\s)$ has an
analytic continuation to $\UU\!\setminus\!\{0\}$ (still denoted with $\F$).
Starting from the Laurent expansion $\F(\s) = \sum_{k = -\infty}^{+\infty}
\F_k \s^k$, we introduce the \textsl{regular part} $(RP\,\F)(\s) :=
\sum_{k=0}^{+\infty} \F_k \s^k$ and define
\beq \F_{ren} := (RP\,\F)(0)~. \label{renest} \feq
Of course, if $\F$ is regular at $\s = 0$ the defnitions \rref{ren}
\rref{renest} coincide. \parn
In the case of the stress-energy VEV, the prescriptions (i) and (ii)
read, respectively,
\beq \la 0 | \Ti_{\mu \nu}(x) | 0 \ra_{ren} :=
\la 0 | \Tis_{\mu \nu}(x) | 0 \ra \Big|_{\s=0} ~, \label{pri} \feq
\beq \la 0 | \Ti_{\mu \nu}(x) | 0 \ra_{ren} :=
RP \Big|_{\s=0} \la 0 | \Tis_{\mu \nu}(x) | 0 \ra ~. \label{prii} \feq
\vspace{-0.9cm}
\subsection{Conformal and non-conformal parts of the stress-energy VEV.} \label{ConfSubsec}
These are indicated by the superscripts $(\Co)$ and $(\NCo)$, respectively;
they are defined by
\beq \la 0|\Ti_{\mu\nu}|0\ra_{ren} = \la 0|\Ti^{(\Co)}_{\mu\nu}|0\ra_{ren}
+ (\xi\!-\!\xi_d)\,\la 0|\Ti^{(\NCo)}_{\mu\nu}|0\ra_{ren} ~, \label{TRinCo}\feq
where we are considering for the parameter $\xi$ the critical value
\beq \xi_d := {d\!-\!1 \over 4d} ~. \label{xic} \feq
\vspace{-0.9cm}
\subsection{Integral kernels.} If $\BB$ is a linear operator in $L^2(\Om)$,
its integral kernel is the (generalized) function $(\bx,\by) \in
\Om\! \times\!\Om \mapsto \BB(\bx,\by) := \la
\de_{\bx}|\BB\,\de_{\by}\raL$ ($\de_\bx$ is the Dirac delta at
$\bx$). The trace of $\BB$, assuming it exists, fulfills $\Tr \BB
= \int_\Om d\bx\,\BB(\bx,\bx)$\,. \parn In the following
subsections $\AA$ is a strictly positive selfadjoint operator in
$L^2(\Om)$, with a complete orthonormal set of eigenfunctions as
in Eq. \rref{eigenf}. In typical applications, $\AA$ is the
operator \rref{defaa}. \vspace{-0.4cm}
\subsection{The Dirichlet kernel and its relations with the stress-energy VEV.}\label{DirT}
For (suitable) $s \in \complessi$, the $s$-th Dirichlet kernel of $\AA$ is
\beq \Dir_s(\bx, \by) := \AA^{-s}(\bx,\by) = \int_\KK {dk \over \om_k^{2s}}\;
\Fk(\bx) \Fkc(\by) ~. \label{eqkerdi} \feq
If $\AA = -\Delta + V$ (with $V$ a smooth potential) is strictly positive, the map
$\Dir_s(~,~):\Om \times \Om \to \complessi$, $(\bx,\by) \mapsto \Dir_s(\bx,\by)$
is continuous along with all its partial derivatives up to order $j \in \naturali$,
for all $s \in \complessi$ with $\Re s > d/2+j/2$\,. \salto
Recalling Eq. \rref{tiquans}, the regularized stress-energy VEV can be
expressed as follows:
\beq {~}\hspace{-0.5cm} \la 0 | \Tis_{0 0}(\bx) | 0 \ra \! = \!\mm^\s
\!\l[\!\l(\!\frac{1}{4}\!+\!\xi\!\r)\!\Dir_{{\s - 1\over 2}}(\bx,\by)
\!+\!\l(\!\frac{1}{4}\!-\!\xi\!\r)\!
(\partial^{x^\ell}\!\partial_{y^\ell}\!+\!V(\bx))
\Dir_{{\s + 1 \over 2}}(\bx,\by)\r]_{\by = \bx}\!, \label{Tidir00} \feq
\beq \la 0 | \Tis_{0 j}(\bx) | 0 \ra =
\la 0 | \Tis_{j 0}(\bx) | 0 \ra = 0 ~, \label{Tidiri0} \feq
\begin{equation}\begin{split}
& {~}\hspace{3.8cm} \la 0 | \Tis_{i j}(\bx) | 0 \ra =
\la 0 | \Tis_{j i}(\bx) | 0 \ra = \\
& = \mm^\s \!\l[\!\Big({1\over 4} - \xi\Big) \de_{i j}
\Big(\!\Dir_{{\s - 1 \over 2}}(\bx,\by) -
(\partial^{\,x^\ell}\!\partial_{y^\ell}\!+\!V(\bx))
\Dir_{{\s + 1 \over 2}}(\bx,\by) \Big) \, + \r. \\
& \hspace{5cm} \l. + \l(\!\Big({1\over 2} - \xi\Big)\partial_{x^i y^j}
- \xi\,\partial_{x^i x^j}\!\r)\!
\Dir_{{\s + 1 \over 2}}(\bx,\by) \r]_{\by = \bx}
\label{Tidirij}
\end{split}\end{equation}
($\la 0|\Tis_{\mu \nu}(\bx)|0\ra$ is short for $\la 0|\Tis_{\mu\nu}(t,\bx)|0\ra$;
indeed, the VEV does not depend on $t$); of course, the map $\Om \to \complessi$,
$\bx \mapsto \la 0|\Tis_{\mu \nu}(\bx)|0\ra$ possesses the same regularity as the
functions $\bx \in \Om \mapsto \Dir_{\s\pm 1 \over 2}(\bx,\bx),
\partial_{zw}\Dir_{\s+1 \over 2}(\bx,\bx)$ ($z,w$ any two spatial variables);
so, due to the previously mentioned results, $\bx \mapsto \la
0|\Tis_{\mu \nu}(\bx)|0\ra$ is continuous for $\Re \s > d+1$.
\parn
The above framework relates the renormalized stress-energy VEV
$\la 0|\Ti_{\mu \nu}(\bx)|0\ra_{ren} := RP|_{\s = 0}\la
0|\Tis_{\mu \nu}(\bx)|0\ra$ to the analytic continuation at $\s =
0$ of the maps $\s \mapsto \mm^\s \Dir_{\s \pm 1 \over
2}(\bx,\by)$ and of their derivatives. \parn In the sequel we will
also consider the total energy VEV and express it in terms of the
trace $\Tr\AA^{-s}$, fulfilling \beq \Tr \AA^{-s} = \int_\Om d\bx
\; \Dir_s(\bx, \bx)~. \label{TrAs} \feq \vspace{-0.9cm}
\subsection{The heat kernel.} For $\t \in [0,+\infty)$, this is given by
\beq K(\t\,;\bx,\by) := e^{-\t\AA}(\bx,\by) =
\int_\KK dk\;e^{- \t\,\om_k^2}\,\Fk(\bx)\Fkc(\by) ~. \label{eqheat} \feq
If $\AA = -\Delta + V$ ($V$ smooth) is strictly positive, the map $K(\t\,;~,~):
\Om \times \Om \to \complessi$, $(\bx,\by) \mapsto K(\t\,;\bx,\by)$ is continuous
along with all its partial derivatives of any order, for all $\t > 0$\,. \salto
The \textsl{heat trace} (assuming it to exists) is
\beq K(\t) := \Tr e^{-\t\AA} = \int_\Om d\bx\;K(\t\,;\bx,\bx) ~.
\label{KTr} \feq
\vspace{-0.9cm}
\subsection{The Dirichlet kernel as Mellin transform of the heat kernel.}
For suitable values of $s\!\in\!\complessi$ (see Part I), there holds
\beq \Dir_s(\bx,\by) = {1 \over \Ga(s)} \int_0^{+\infty}
\!d\t \; \t^{s-1}\,K(\t\,;\bx,\by) ~. \label{DirHeat} \feq
Similar results hold for $\Tr \AA^{-s}$; for example, using the
heat trace $K(\t)$ of Eq. \rref{KTr}, we obtain
\beq \Tr \AA^{-s} = {1 \over \Ga(s)}\int_0^{+\infty}
\!d\t \; \t^{s-1}\,K(\t) ~. \label{DirHeatTr} \feq
\subsection{Analytic continuation of Mellin transforms via integration by parts.}\label{contiparts}
Let $\ff:(0,+\infty) \to \complessi$ be a function of the form
\beq \ff(\t) = {1 \over \t^\rho}\,\hh(\t) \label{ff} \feq for some
$\rho \in \complessi$ and some smooth function $\hh:[0,+\infty)
\to \complessi$, vanishing exponentially for $\t \to +\infty$.
Consider the Mellin transform of $\ff$, i.e. \beq \Mel(\si) :=
\int_0^{+\infty}\!\!d\t\;\t^{\si-1}\,\ff(\t) ~; \label{Mel}\feq
this is an analytic function of $\si$ for $\Re \si > \Re \rho$.
Integrating the above equation by parts $n$ times (for any $n \in
\{1,2,3,...\}$) we obtain the relation \beq \Mel(\si) = {(-1)^n
\over (\si\!-\!\rho) ... (\si\!-\!\rho\!+\!n\!-\!1)}
\int_0^{+\infty} \!\!d\t \; \t^{\si-\rho+n-1}\,{d^n \hh \over d
\t^n}(\t) ~, \label{MelCon} \feq giving the analytic continuation
of $\Mel(\si)$ to a meromorphic function of $\si$ in the region
$\{\Re \si > \Re \rho - n\}$, possibly with simple poles at $\si
\in \{\rho\,,\,\rho - 1\,,\,...\,,\,\rho - n + 1\}$. \parn
The above results can be employed to obtain the analytic continuations
of the regularized stress-energy VEV $\la 0|\Tis_{\mu\nu}(\bx)|0\ra$
(treating $\bx\in\Om$ as a fixed parameter) and of the trace $\Tr \AA^{-s}$;
for example, assuming the heat trace $K(\t)$ to have the form
\beq K(\t) = {1 \over
\t^p}\;H(\t) ~, \label{espkTr} \feq for some $p \in \reali$ and
some smooth function $H: [0,+\infty) \to \reali$, vanishing
rapidly for $\t \to +\infty$, starting from Eq. \rref{DirHeatTr}
we obtain (for $n \in \{1,2,3,...\}$) \beq \Tr \AA^{-s} = {(-1)^n
\over \Ga(s)(s\!-\!p)...(s\!-\!p\!+\!n\!-\!1)} \int_0^{+\infty}
\!\!d\t \;\t^{s-p+n-1}\, {d^n H \over d \t^n}(\t) ~.
\label{HeatConTr} \feq
\vspace{-0.7cm}
\subsection{The case of product domains. Factorization of the heat kernel.} \label{prodDom}
Let $\AA := -\Delta + V$ and consider the case where \beq \Om =
\Om_1 \times \Om_2 \ni \bx = (\bx_1, \bx_2)\,, \by = (\by_1,\by_2)
~, \label{omfact} \feq \beq V(\bx) = V_1(\bx_1) + V_2(\bx_2) \feq
($\Om_a\!\subset\!\reali^{d_a}$ is an open subset, for $a \in
\{1,2\}$; $d_1\!+\!d_2=d$); assume the boundary conditions on
$\partial \Om$ to arise from suitable boundary conditions
prescribed separately on $\partial \Om_1$ and $\partial \Om_2$ so
that, for $a=1,2$, the operators \beq \AA_a := - \Delta_a +
V(\bx_a) \feq (with $\Delta_a$ the Laplacian on $\Om_a$) are
selfadjoint and strictly positive in $L^2(\Om_a)$. Then, the
Hilbert space $L^2(\Om)$ and the operator $\AA$ can be represented
as \beq L^2(\Om) = L^2(\Om_1) \otimes L^2(\Om_2) ~, \qquad \AA =
\AA_1 \otimes \uno + \uno \otimes \AA_2 ~. \label{prodOmAA}\feq
This implies, amongst else, that the heat kernels $K(\t\,;\bx,\by)
:= e^{\t \AA}(\bx, \by)$, $K_a(\t\,;\bx_a,\by_a)$ $:= e^{\t
\AA_a}(\bx_a,\by_a)$ ($a = 1,2$) are related by \beq
K(\t\,;\bx,\by) = K_1(\t\,;\bx_1,\by_1)\,K_2(\t\,;\bx_1,\bx_2) ~.
\label{prodK} \feq Similarly, writing $K(\t)$, $K_a(\t)$ ($a
\in\{1,2\}$) for the heat traces of $\AA$ and $\AA_a$
($a\in\{1,2\}$), respectively, we have \beq K(\t) =
K_1(\t)\,K_2(\t) ~. \label{prodKTr} \feq Let us briefly recall
that a special subcase of the present framework is the case of a
\textsl{slab}, i.e., \beq \Om = \Om_1 \times \reali^{d_2}\ni \bx =
(\bx_1, \bx_2)\,, \by = (\by_1,\by_2) ~, \qquad V(\bx) = V(\bx_1)
\feq with $\Om_1 \subset \reali^{d_1}$ an open subset
($d_1\!+\!d_2 = d$);
see Part I for more details on this topic.\\
$\phantom{a}$ \vspace{-0.9cm}
\subsection{The total energy.}\label{TotEnSub} The \textsl{zeta-regularized
total energy} is
\beq \EE^\s := \int_\Om d\bx\; \la 0|\Tis_{00}(\bx)|0\ra = E^\s + B^\s ~; \label{EEtot}\feq
the second equality is proved after defining the \textsl{regularized bulk}
and \textsl{boundary energies}, which are
\beq E^\s := {\mm^\s\! \over 2} \int_\Om d\bx\;\Dir_{{\s - 1\over 2}}(\bx,\bx)
= {\mm^\s\! \over 2}\; \Tr\,\AA^{{1 -\s \over 2}} ~, \label{defEs}\feq
\beq B^\s := \mm^\s \l(\!{1 \over 4}-\xi\!\r)\! \int_{\partial \Om}\!
d a(\bx)\,\l.{\partial \over \partial n_{\by}}\,
\Dir_{{\s + 1\over 2}}(\bx,\by)\r|_{\by = \bx} ~. \label{defBs} \feq
If $\Om$ is unbounded, the integral over $\partial\Om$ in Eq. \rref{defBs}
has to be intended as $\lim_{\ell \to + \infty} \int_{\partial \Om_\ell} da$,
where $(\Om_\ell)_{\ell = 0,1,2,...}$ is a sequence of bounded subdomains
with $\Om_\ell \subset \Om_{\ell+1}$ for $\ell \in \{0,1,2,..\}$, and
$\cup_{\ell=0}^{+\infty}\Om_\ell = \Om$. \parn
Assuming $E^\s$ and $B^\s$ (see Eq.s \rref{defEs} \rref{defBs}) to be finite
and analytic for suitable $\s \in \complessi$, we define the renormalized
energies by the generalized (or restricted) zeta approach; for example, we put
\beq E^{ren} := RP \Big|_{\s=0} E^\s \qquad \l(\mbox{or } \;
E^{ren} := E^\s \Big|_{\s=0}\,\r)\,. \label{ERegnGen} \feq
\subsection{Curvilinear coordinates.} \label{curvSubsec}
In some applications it is natural to employ on $\Om$ some set of
curvilinear coordinates $(q^i)_{i = 1,...,d} \equiv \bq$, inducing
a set of spacetime coordinates $q \equiv (q^\mu) \equiv (t,\bq)$;
the spatial and space-time line elements are, respectively,
\begin{equation}\begin{split}
& \hspace{1.2cm} d\ell^2 = a_{i j}(\bq)dq^i dq^j ~; \qquad
ds^2 = - dt^2\!+\!d\ell^2 = g_{\mu \nu}(q)\, dq^\mu dq^\nu ~, \\
& g_{00} := -1 ~, ~\quad g_{i 0} = g_{0 i} := 0~, ~\quad
g_{i j}(q) := a_{i j}(\bq) \qquad \mbox{for $i,j \in \{1,...,d\}$}~. \label{dsq}
\end{split}\end{equation}
The analogue of Eq. \rref{tiquans} in the coordinate system $(q^\mu)$ is
\beq \Tis_{\mu \nu} := (1\!-\!2\xi)\partial_\mu \Fis\!\circ\partial_\nu \Fis\!
- \!\l({1\over 2}\!-\!2\xi\!\r)\!\eta_{\mu\nu}\!
\l(\!\partial^\lam\Fis\partial_\lam \Fis\!+\!V(\Fis)^2\r)
- 2 \xi \, \Fis\!\circ \nabla_{\!\mu \nu} \Fis \label{tiquansq} \feq
($\nabla_{\mu}$ is the covariant derivative induced by the metric
\rref{dsq}). For any scalar function $f$ there hold ($\ga^{k}_{ij}$
are the Christoffel symbols for the spatial metric $(a_{ij}(\bq))$,
$D_i$ the corresponding covariant derivative)
\begin{equation}\begin{split}
& \hspace{0.6cm} \nabla_{\!\mu} f = \partial_\mu f ~, \qquad
\nabla_{ij} f = D_{ij}f = \partial_{ij} f - \ga^{k}_{ij} \partial_k f ~,\\
& \nabla_{0 i} f = \partial_0(\partial_{i}f) =  \partial_i(\partial_0 f) =
\nabla_{i 0}f ~, \qquad \nabla_{0 0} f = \partial_{0 0} f ~.
\label{compu}
\end{split}\end{equation}
In the present paper we often work in a curvilinear coordinate system;
more precisely we use a set of (rescaled) spherical coordinates, fitting
the symmetries of the configuration under analysis. Most of the results
of the previous subsections are readily rephrased in this framework.
\section{The case of a harmonic potential} \label{secHarm}
\subsection{Introducing the problem.} We consider the case of a massless
scalar field on $\reali^d$ in presence of a classical isotropic harmonic
potential. More precisely, we assume
\beq \Om := \reali^d, \qquad V(\bx) := k^4\,|\bx|^2 ~, \label{harm} \feq
where $k>0$ and $|\bx| := \sqrt{(x^1)^2 + ... + (x^d)^2}$\,; the
constant $k$ is, dimensionally, a mass (or an inverse length) like
the parameter $\mm$ employed for the field regularization
\rref{Fis}. \parn
All the considerations reported in the following could be generalized,
with some calculational effort, to the anisotropic case $V(\bx) =
\sum_{i=1}^d (k_i)^4(x^i)^2$ where $k_i \geqs 0$ for all $i \in \{1,...,d\}$\,,
also including slab cases where some of the $k_i$ vanish; a further
variation would concern a massive field in one of the above mentioned
configurations, so that, e.g., in place of \rref{harm} we would have
$V(\bx) := m^2 + k^4 |\bx|^2$\,. None of these generalizations will be
considered, for the sake of computational simplicity.
\subsection{The heat kernel.} \label{heatHarm}
Even though the configuration \rref{harm} is patently spherically
symmetric, we choose to postpone the use of spherical coordinates to
the next subsection; here we work in standard Cartesian coordinates
$(x^i)_{i = 1,...,d}$, for reasons that will soon become apparent. \parn
First of all, notice that \rref{harm} is a configuration of the product
type considered in subsection \ref{prodDom}; in particular, with a
trivial generalization of the results therein, we have for the Hilbert
space of the system and for the fundamental operator $\AA := - \Delta + V$,
respectively, the following representations (compare with Eq. \rref{prodOmAA}):
\beq L^2(\reali^d) = \bigotimes_{i=1}^d L^2(\reali)~, \feq
\beq \AA = \AA_1 \otimes \uno \otimes ... \otimes \uno + ... +
\uno \otimes ... \otimes \uno \otimes \AA_1~, \quad\;
\AA_1 := - {d^2 \over d x^2} + k^4 x^2 ~~ \mbox{on $L^2(\reali)$} ~. \feq
In this situation, the heat kernel of $\AA$ is given by (compare with
Eq. \rref{prodK})
\beq K(\t\,;\bx,\by) = \prod_{i=1}^d K_1(\t\,; x^i, y^i) \feq
where $K_1$ is the heat kernel of $\AA_1$; the latter is the familiar
\textsl{Mehler kernel}, that is
\beq K_1(\t\,;x,y) = {k \over \sqrt{2 \pi \sinh(2 k^2 \t)}}\,
\exp\l[-k^2 \l({x^2\!+y^2 \over 2 \tanh(2 k^2\t)}
- {x\,y \over \sinh(2k^2\t)} \r) \r] \label{K1Harm}\feq
(see \cite{Calin}, using the equations therein with the replacement
$(t,x,y) \to (k^2 t,kx,ky)$). The last two relations imply
(with $\bx \cdot \by := \sum_{i=1}^d x^i\, y^i$)
\beq K(\t\,;\bx,\by) = \l({k \over \sqrt{2 \pi \sinh(2 k^2 \t)}}\r)^{\!\!d}
\exp\!\l[-k^2\!\l({|\bx|^2\!+|\by|^2 \over 2 \tanh(2 k^2 \t)}
-{\bx \cdot \by \over \sinh(2k^2 \t)} \r)\!\r]\,. \label{equak} \feq
A similar analysis can be performed for the heat trace. Firstly note
that (compare with Eq. \rref{prodKTr})
\beq K(\t) = \prod_{i=1}^d K_1(\t) = \Big(K_1(\t)\Big)^d \feq
where $K_1(\t)$ indicates the trace for the reduced one-dimensional
problem; for the latter quantity we can easily deduce the expression
({\footnote{In order to derive Eq. \rref{K1HarmTr} we can proceed, for
example, as explained hereafter. The second equality in Eq. \rref{KTr}
and the explicit expression \rref{K1Harm} for the one-dimensional heat
kernel (along with some trivial trigonometric identities) allow us to infer
$$ K_1(\t) = {k \over \sqrt{2 \pi \sinh(2 k^2 \t)}} \int_\reali dx\;
e^{-k^2 x^2 \tanh(k^2\t)} ~; $$
then, Eq. \rref{K1HarmTr} follows evaluating the above elementary
Gaussian integral.}})
\beq K_1(\t) = {1 \over 2 \sinh(k^2\t)} ~. \label{K1HarmTr}\feq
Summing up, for the heat trace of the complete $d$-dimensional problem
we have
\beq K(\t) = \l({1 \over 2 \sinh(k^2\t)}\r)^{\!\!d} ~. \label{equakTr} \feq
In the following subsections we explain how to construct the analytic
continuation of the zeta-regularized VEV of the stress-energy tensor,
using the expression \rref{equak} for the heat kernel and the integral
representation \rref{DirHeat} for the Dirichlet kernel.
Moreover, we show how to determine asymptotic expressions for the
renormalized stress-energy VEV in the limit of small and large $|\bx|$,
respectively. Moving along the same lines, we also derive the renormalized
VEV of the total energy; to this purpose, we consider the representation
\rref{DirHeatTr} for the trace $\Tr\AA^{-s}$ and the expression
\rref{equakTr} for the heat trace.
$\phantom{a}$ \vspace{-0.9cm} \\
\subsection{Spherical coordinates.} As anticipated in subsection
\ref{heatHarm}, we now pass to a set of curvilinear coordinates which
best fit the symmetries of the problem (see subsection \ref{curvSubsec});
namely, we introduce the spherical ``$k$-rescaled'' coordinates
\beq \bx \mapsto \bq(\bx) \equiv (r(\bx),\te_1(\bx),...,\te_{d-2}(\bx),
\te_{d-1}(\bx)) \in (0,+\infty) \times (0,\pi) \times ... \times (0,\pi)
\times (0,2\pi) \feq
whose inverse map $\bq \mapsto \bx(\bq)$ is described by the equations
\begin{equation}\begin{split}
k\,x^1 & = r \cos(\te_1) ~, \\
k\,x^2 & = r \sin(\te_1)\cos(\te_2) ~, \\
& \;\;\vdots \\
k\,x^{d-1} & = r \sin(\te_1)...\sin(\te_{d-2})\cos(\te_{d-1}) ~, \\
k\,x^d & = r \sin(\te_1)...\sin(\te_{d-1}) ~. \label{sphCoo}
\end{split}\end{equation}
Needless to say, for $d = 1$ (when the system is invariant under the
parity symmetry $x^1 \to - x^1$) we set
\beq r := k\,|x^1| ~. \label{ri1}\feq
Let us stress that, for any spatial dimension $d$, we have
\beq r = k\,|\bx| ~; \feq
thus, the coordinate $r$ is an adimensional radius.
In order to avoid cumbersome notations, given a function $\Om \to Y$,
$\bx \mapsto f(\bx)$ (with $Y$ any set), we indicate the composition
$\bq \mapsto f(\bx(\bq))$ as $\bq \mapsto f(\bq)$\,; we will use
similar conventions for functions on $\Om \times \Om$ (see, e.g.,
what follows). \parn
Let us now consider the Dirichlet function $\Dir_s(\bx,\by)$ and the
heat kernel $K(\t\,;\bx,\by)$; let
\beq \bq = (r,\te_1,...,\te_{d-1}) \equiv (r,\boma{\te}) ~, \qquad
\bp = (r',\te'_1,...,\te'_{d-1}) \equiv (r,\boma{\te}') \label{bqbp} \feq
and put
\beq \tau := k^2\t \;\in(0,+\infty) ~. \label{ttau}\feq
In the sequel we write $\Dir_s(\bq,\bp)$ and $K(\tau\,;\bq,\bp)$, respectively,
for the Dirichlet and heat kernels at two points $\bx,\by$ of (rescaled)
spherical coordinates $\bq,\bp$, and with $\tau$ related to $\t$ by Eq. \rref{ttau}.
Eq. \rref{equak} implies
\beq K(\tau\,;\bq,\bp) = \l({k \over \sqrt{2 \pi \sinh(2 \tau)}}\r)^{\!\!d}
\exp\!\l[-\!\l({r^2\!+{r'}^{\,2} \over 2 \tanh(2 \tau)}
-{r\,r'\,S(\boma{\te})S(\boma{\te}') \over \sinh(2 \tau)} \r)\r] \label{equakSph} \feq
where $S(\boma{\te})$ and $S(\boma{\te}')$ are the products of cosines
and sines of the angular coordinates $(\te_1,...,\te_{d_1})$ and
$(\te'_1,...,\te'_{d_1})$ of Eq. \rref{bqbp}, corresponding to the
scalar product $\bx \cdot \by$. \parn
From Eq. \rref{DirHeat} it follows that
\beq \Dir_s(\bq,\bp) = {k^{-2s} \over \Ga(s)} \int_0^{+\infty}
\!d\tau \; \tau^{s-1}\,K(\tau\,;\bq,\bp) ~; \label{DirHeatSph} \feq
substituting this relation into the analogues of Eq.s (\ref{Tidir00}-\ref{Tidirij})
for curvilinear coordinates (see subsection \ref{curvSubsec}), we get
\beq {~}\hspace{-0.4cm}\la 0|\Tis_{\mu\nu}(\bq)|0\ra\!= {k^{d+1} \over \Ga({\s+1 \over 2})}
\l({\mm \over k}\r)^{\!\!\s}\!\! \int_0^{+\infty}\hspace{-0.41cm}d\tau\;\tau^{{\s-d-3 \over 2}}\,
\HH^{(\s)}_{\mu \nu}(\tau\,;\bq)~~ (\mu,\nu\!\in\!\{0,r,\te_1,...,\te_{d-1}\})
\label{TisHarmPol} \feq
where the coefficients $\HH^{(\s)}_{\mu\nu}(\tau\,;\bq)$ are as follows
(for $i,j,h,\ell\in \{\rho,\te_1,...,\te_{d-1}\}$; here $D$ is the
spatial covariant derivative)
\begin{equation}\begin{split}
& \hspace{5.5cm} \HH^{(\s)}_{00}(\tau\,;\bq) := \\
& \l({\tau \over k^2}\r)^{\!\!d/2}\!
\bigg[\!\!\l(\!\frac{1}{4}\!+\!\xi\!\r)\!\!\l(\!{\s\!-\!1 \over 2}\!\r)\!
+\!\l(\!\frac{1}{4}\!-\!\xi\!\r)\!\tau\,
\Big(a^{h\ell}(\bq)D_{q^h p^\ell} + r^2\Big)\!\bigg]\bigg|_{\bp = \bq}
K(\tau\,;\bq,\bp) ~, \label{H00Gen}
\end{split}\end{equation}
\beq \HH^{(\s)}_{0i}(\tau\,;\bq) = \HH^{(\s)}_{i0}(\tau\,;\bq) := 0 ~, \feq
\begin{equation}\begin{split}
& \hspace{4.7cm} \HH^{(\s)}_{ij}(\tau\,;\bq) = \HH^{(\s)}_{ji}(\tau\,;\bq) := \\
& \l({\tau \over k^2}\r)^{\!\!d/2}\!\bigg[\!\Big({1\over 4} - \xi\Big)a_{i j}(\bq)\!
\l({\s\!-\!1 \over 2}\,- \tau\,\Big(a^{h\ell}(\bq)D_{q^h p^\ell} + r^2\Big)\!\r) + \\
& \hspace{4.1cm} + \l({\tau \over k^2}\r)\!
\l(\!\Big({1\over 2} - \xi\Big)D_{q^i p^j}- \xi\,D_{q^i q^j}\!\r)\!\bigg]
\bigg|_{\bp = \bq}K(\tau\,;\bq,\bp) ~. \label{HijGen}
\end{split}\end{equation}
Here and in the reaminder of the paper, we are implicitly understanding
the dependence on the parameter $\xi$ for simplicity of notation: so,
$\HH^{(\s)}_{\mu\nu}(\tau\,;\bq)$ stands for $\HH^{(\s)}_{\mu\nu}(\tau\,;\bq\,;\xi)$. \parn
In the following subsection we point out some notable properties of the
coefficients $\HH^{(\s)}_{\mu\nu}(\tau\,;\bq)$; these properties will be
employed in the following subsection \ref{ACH} in order to evaluate the
renormalized stress-energy VEV.
\vspace{-0.4cm}
\subsection{Properties of the coefficients $\boma{\HH^{(\s)}_{\mu \nu}}$.}\label{propHH}
Consider the explicit expressions (\ref{H00Gen}-\ref{HijGen}) for these
coefficients. Using as well Eq. \rref{equakSph} for the heat kernel
expressed in rescaled, spherical coordinates, one can prove the following
facts. \vspace{0.1cm}\\
i) For fixed $\tau,\bq$ and any $\mu,\nu \in \{0,r,\te_1,...,\te_{d-1}\}$,
the map $\s \mapsto \HH^{(\s)}_{\mu \nu}(\tau\,;\bq)$ is affine. \vspace{0.1cm}\\
ii) For fixed $\bq$ and any $\s \in \complessi$, $\mu,\nu \in
\{0,r,\te_1,...,\te_{d-1}\}$, the map $\tau \mapsto \HH^{(\s)}_{\mu\nu}(\tau\,;\bq)$
is smooth on $[0,+\infty)$ and exponentially vanishing for $\tau \to +\infty$. \vspace{0.1cm}\\
iii) The final expressions for the coefficients $\HH^{(\s)}_{\mu\nu}$ do
not depend on the parameter $k$ (even though it appears in the right-hand
sides of Eq.s (\ref{H00Gen}-\ref{HijGen})). \vspace{0.1cm}\\
iv) There hold
\begin{equation}\begin{split}
& \hspace{4cm} \HH^{(\s)}_{\mu \nu} = 0 \quad \mbox{for $\mu \neq \nu$}~, \\
& \HH^{(\s)}_{\te_{d-1} \te_{d-1}} = \sin^2(\te_{d-2})\,\HH^{(\s)}_{\te_{d-2} \te_{d-2}}
= ... = \sin^2(\te_{d-2})...\sin^2(\te_1)\,\HH^{(\s)}_{\te_1 \te_1} ~. \label{HHprel}
\end{split}\end{equation}
v) For $\mu = \nu \in \{0,r,\te_1\}$, there hold
\beq \HH^{(\s)}_{\mu \nu}(\tau\,;\bq) =
e^{-r^2 \tanh \tau} \MM^{(\s)}_{\mu \nu}(\tau;r) \label{antic} \feq
where $\MM^{(\s)}_{\mu \nu}(\tau;r)$ is a polynomial in $r,\s$ of
degree $1$ in both these variables, with coefficients depending
smoothly on $\tau$. \salto
All the above statements i)-v) could be proved for arbitrary $d$, but we
will limit ourself to check them in the cases $d \in \{1,2,3\}$ considered
hereafter, for which the explicit expressions of all coefficients
$\HH^{(\s)}_{\mu \nu}(\tau\,;\bq)$ ($\mu = \nu \in \{0,r,\te_1\}$) will
be given: see the subsequent subsections \ref{Harm1}, \ref{Harm2} and
\ref{Harm3}, respectively. \parn
Before moving on, let us make a few remarks. First, consider the integral
representation \rref{TisHarmPol} for the regularized stress-energy VEV;
in consequence of point iii), the latter VEV only depends on the parameter
$k$ through the multiplicative coefficient $k^{d+1} (\mm/k)^\s$ in front
of the integral in the cited equation. Next, we note that Eq. \rref{HHprel}
of point iv) indicates that the VEV $\la 0|\Tis_{\mu\nu}(\bq)|0\ra$ is
diagonal and that the only independent components are those with
$\mu = \nu \in \{0,r,\te_1\}$; finally, these components only depend on
the radial coordinate $r$ (and not on the angular ones $\{\te_1,...,\te_{d-1}\}$),
in accordance with the spherical symmetry of the configuration under analysis.
\vspace{-0.4cm}
\subsection{Analytic continuation.}\label{ACH} Let us move on to determine
the analytic continuation of the regularized VEV \rref{TisHarmPol}. To this
purpose, we refer to the framework of subsection \ref{contiparts}; using
Eq.s (\ref{ff}-\ref{MelCon}) with $\HH = \HH^{(\s)}_{\mu\nu}(~;\bx)$,
$\rho = (d+3-\s)/2$, we obtain
\begin{equation}\begin{split}
& \hspace{5.3cm} \la 0|\Tis_{\mu\nu}(\bq)|0\ra = \\
& {k^{d+1} \over \Ga({\s+1 \over 2})}\,
{(-1)^n \over ({\s-d-3 \over 2}\!+\!1)...({\s-d-3 \over 2}\!+\!n)}
\l({\mm \over k}\r)^{\!\!\s}\!\!\int_0^{+\infty}\!\! d\tau\;
\tau^{{\s-d-3 \over 2}+n}\,\partial_{\tau}^{n}\HH^{(\s)}_{\mu \nu}(\tau;\bq) ~.
\label{TiHarmAC}
\end{split}\end{equation}
Notice that, due to the features of $\HH^{(\s)}_{\mu \nu}(\tau;\bq)$ pointed out
in the previous subsection, the integral in the above expression converges for
\beq \Re \s > d + 1 - 2n ~, \label{Resn}\feq
so that Eq. \rref{TiHarmAC} yields the required analytic continuation of
$\la 0|\Tis_{\mu\nu}(\bq)|0\ra$ to the very same region. Since we are
interested in evaluating (the regular part of) this analytic continuation
in $\s = 0$, we choose $n$ so that Eq. \rref{Resn} holds for $\s = 0$, i.e.
({\footnote{As a matter of fact, for computational simplicity, we will
always choose the smaller $n \in \{0,1,2,...\}$ fulfilling Eq. \rref{nMin}.}})
\beq n > {d + 1 \over 2} ~. \label{nMin} \feq
Following Eq. \rref{renest}, in general we define
\beq \la 0|\Ti_{\mu\nu}(\bq)|0\ra_{ren} :=
RP\Big|_{\s = 0} \la 0|\Tis_{\mu\nu}(\bq)|0\ra ~. \label{THarmRen} \feq
For $n$ as in Eq. \rref{nMin}, consider the expression on the right-hand
side of Eq. \rref{TiHarmAC}; for any even spatial dimension $d$ this
expression is regular in $\s = 0$, so that we can apply the zeta
regularization in its restricted version to obtain the renormalized
stress-energy VEV. On the other hand, for odd $d$ we must resort to the
extended version of the zeta technique, since the function under analysis
has a simple pole in $\s = 0$\,. \parn
Because of the pole singularity, the procedure of evaluating the regular
part of Eq. \rref{TiHarmAC} in $\s = 0$ in the case of odd $d$ implies
the appearance of a logarithmic term in $\tau$ in the integrand
({\footnote{In fact
$$ \tau^{\s/2} = e^{{\s \over 2}\ln\tau} = 1 + {\s \over 2}\,\ln\tau + O(\s^2)
\qquad \mbox{for $\s \to 0$} ~. $$
The logarithmic term proportional to $\s$ is not relevant in the case
of even $d$, where analytic continuation exists up to $\s=0$ and it is
simply obtained setting $\s=0$ in \rref{TiHarmAC}. On the contrary, for
odd $d$ we must take the regular part at $\s=0$ of the expression
\rref{TiHarmAC} and the above term ${\s \over 2}\,\ln\tau$ contributes to
it, since it is multiplied by a term proportional to $1/\s$ that comes
from the $\s \to 0$ expansion of the factor ${1/({\s-d-3 \over 2}\!+\!1)
...({\s-d-3 \over 2}\!+\!n)}$ in \rref{TiHarmAC}.
Also recall the statement in point i) of subsection \ref{propHH}.
}}). Simple but rather lenghty computations give the following results,
for $d$ either odd or even and $\mu,\nu \in \{0,r,\te_1\}$:
\beq \la 0|\Ti_{\mu\nu}(\bq)|0\ra_{ren} = k^{d+1}\Big(T^{(0)}_{\mu\nu}(r)
+ \Mk \,T^{(1)}_{\mu\nu}(r)\Big) ~, \quad \mbox{where} \label{TiRenHarm} \feq
\begin{equation*}\begin{split}
& \hspace{0.9cm} T^{(0)}_{\mu\nu}(r) := \int_0^{+\infty}\!\!\!d\tau\;
\tau^{n-{d+3 \over 2}}\;e^{-r^2\tanh\tau} \Big[\PP^{(0)}_{\mu\nu}(\tau\,;r)
+ \ln\tau\;\PP^{(1)}_{\mu\nu}(\tau\,;r)\Big] ~,\\
& T^{(1)}_{\mu\nu}(r) := \int_0^{+\infty}\!\!\!d\tau\;
\tau^{n-{d+3 \over 2}}\;e^{- r^2\tanh\tau}\;\PP^{(1)}_{\mu\nu}(\tau\,;r) ~,
\qquad \Mk := \ga_{EM}\!+\!2\ln\!\l(\!{2\mm \over k}\r)
\end{split}\end{equation*}
($\ga_{EM} \simeq 0.577$ is the Euler-Mascheroni constant). In the above
$\PP^{(0)}_{\mu\nu}(\tau\,;r)$ and $\PP^{(1)}_{\mu\nu}(\tau\,;r)$ are
suitable functions determined by $\HH^{(\s)}_{\mu \nu}(\tau;\bq)$; these
functions are in fact polynomials in $r^2$ of the form
\beq \PP^{(a)}_{\mu\nu}(\tau\,;r) = \sum_{i = 0}^{n+1}
p^{(a)}_{i,\mu\nu}(\tau)\,r^{2i} \qquad (a \in \{0,1\}) ~, \label{PPPol} \feq
for some smooth functions $p^{(0)}_{i,\mu\nu},p^{(1)}_{i,\mu\nu}:[0,+\infty)
\to \reali$ ($i \in \{0,...,n+1\}$). Let us stress that
\beq \PP^{(1)}_{\mu\nu}(\tau\,;r) = 0~, ~~
T^{(1)}_{\mu\nu}(r) = 0 \qquad \mbox{for $d$ even}~, \label{PP1Zer} \feq
a fact corresponding to the previous comments on the logarithmic terms. \parn
Next note that, in consequence of the remarks of subsection \ref{propHH},
the renormalized VEV $\la 0|\Ti_{\mu\nu}(\bq)|0\ra_{ren}$ only depends
on the parameter $k$ through the coefficients $k^{d+1}$ and $\Mk$ in the
first equation of \rref{TiRenHarm}. In particular, the functions
$T^{(a)}_{\mu\nu}(r)$ ($a \in \{0,1\}$) introduced therein do not
depend on $k$ and we can evaluate them computing the integrals in
Eq. \rref{TiRenHarm} numerically, for any fixed $r \in (0,+\infty)$. \parn
To conclude, following the considerations of subsection \ref{ConfSubsec}
({\footnote{In particular, we use the same convention as in Part I (see
Eq. (2.24) therein or Eq. \rref{TRinCo} in the present paper):
$$ \mbox{$\Co \equiv$ conformal}~, \qquad \mbox{$\NCo \equiv $ non-conformal} ~. $$
Compare Eq. \rref{TiCNC} with Eq.s (2.25) (2.26) in Part I.}}),
we define the conformal and non-conformal parts of the functions
$r \mapsto T^{(0)}_{\mu\nu}(r), T^{(1)}_{\mu\nu}(r)$ respectively as
\beq T^{(a,\Co)}_{\mu\nu} := T^{(a)}_{\mu\nu}\Big|_{\xi = \xi_d}~,
\quad T^{(a,\NCo)}_{\mu\nu} := {1 \over \xi\!-\!\xi_d}\,
\Big(T^{(a)}_{\mu\nu} - T^{(a,\Co)}_{\mu\nu}\Big) \qquad (a \in \{0,1\}) \label{TiCNC}\feq
($\xi_d$ is the conformal parameter, see Eq. \rref{xic}).
In subsections \ref{Harm1}-\ref{Harm3} we present the functions $\PP^{(a)}_{\mu\nu}$
($a \in \{0,1\}$) and the graphs (obtained via numerical integration) for each one
of the functions in Eq. \rref{TiCNC}, for $d \in \{1,2,3\}$ respectively.
\vspace{-0.4cm}
\subsection{Asymptotics for $\boma{r = k|\bx| \to 0^{+}}$.} \label{smallri}
Hereafter we present a method which we will use to determine the asymptotic
expansion for the renormalized stress-energy VEV in the limit $r = k|\bx| \to 0^+$. \parn
Let us consider a function $F:(0,+\infty)\to\reali$ defined by
\beq F(r) := \int_0^{+\infty}\!\!d\tau\; e^{-r^2 h(\tau)}\,P(\tau\,;r) ~,
\label{FG}\feq
where $h(\tau)$ is a positive bounded function, while $P(\tau\,;r)$ is
a polynomial of degree $N$ in $r^2$; more precisely, we assume that
\beq P(\tau\,;r) = \sum_{i = 0}^{N} p_i(\tau)\,r^{2 i} ~, \label{GPol} \feq
for some integrable functions $p_i : (0,+\infty) \to \reali$
($i\in\{0,...,N\}$). We claim that the Taylor expansion of the exponential
in Eq. \rref{FG} implies an expansion
\beq F(r) = \sum_{i=0}^{N} a_i \,r^{2 i} + R_{N+1}(r)~,
\qquad R_{N+1}(r) = O(r^{2(N+1)})~\mbox{for $r \to 0^+$} ~, \label{FserZer} \feq
where the coefficients $a_i \in \reali$ are defined by
\beq a_i := \sum_{j=0}^i {(-1)^{i-j}\!\over (i\!-\!j)!}\int_{0}^{+\infty}\!\!
d\tau\; p_j(\tau)\,(h(\tau))^{i-j} \qquad (i\in\{0,...,N\}) ~; \label{amn}\feq
more precisely, the remainder term is such that
\beq |R_{N+1}(r)| \leqs C_{N+1} r^{2 (N+1)} \qquad \mbox{for all
$r \in (0,+\infty)$} ~, \label{RCZer}\feq
\beq C_{N+1} := \sum_{i=0}^{N} {1 \over (N\!-\!i\!+\!1)!} \int_0^{+\infty}\!
\!d\tau\;|p_i(\tau)|\,(h(\tau))^{N-i+1} ~. \label{ecn} \feq
See Appendix \ref{AppII} for the derivation of the above results.\salto
Consider the expression \rref{TiRenHarm} for the independent components of
the renormalized stress-energy VEV; recalling that $\PP^{(0)}_{\mu\nu}$
and $\PP^{(1)}_{\mu\nu}$ are both polynomials in $r^2$ of degree $n\!+\!1$
(see Eq. \rref{PPPol}), we note that each component (fixed $\mu,\nu$) has
the form \rref{FG} \rref{GPol} with
\begin{equation}\begin{split}
& \hspace{5cm} h(\tau) = \tanh(\tau) ~, \\
& P(\tau\,;r) = \tau^{n-{d+3 \over 2}} \Big(\PP^{(0)}_{\mu\nu}(\tau\,;r)
+ (\Mk\!+\!\ln\tau)\,\PP^{(1)}_{\mu\nu}(\tau\,;r)\Big)~, \quad N = n+1 ~.
\end{split}\end{equation}
Thus, using the above approach, we can infer the expansion for the
renormalized stress-energy VEV in the limit $r = k|\bx| \to 0^+$;
we defer the explicit results for $d \in\{1,2,3\}$ to Section \ref{Harmd123},
where the integrals in Eq. \rref{amn} are evaluated numerically for the
choices of $h$ and $P$ under analysis.
\vspace{-0.4cm}
\subsection{Asymptotics for $\boma{r = k|\bx| \to + \infty}$.} \label{largeri}
Let us discuss some general methods allowing to determine the
asymptotic behaviour for certain types of integrals; these methods
will be employed to determine the asymptotic expansion of renormalized
stress-energy VEV in the limit $r = k|\bx| \to +\infty$. \parn
Consider a function $F:(0,+\infty)\to\reali$ defined by
\beq F(r) := \int_{0}^{1} dv\; e^{-r^2 v}\,v^\aa\,Q(v;r) \qquad
(\aa > -1) ~, \label{FGInf}\feq
where $Q(v;r)$ is a polynomial in $r^2$ of the form
\begin{equation}\begin{split}
& \hspace{0.6cm} Q(v;r) = Q^{(0)}(v;r) + Q^{(1)}(v;r)\ln v ~, \\
& Q^{(a)}(v;r) = \sum_{i = 0}^{N} q^{(a)}_i(v)\,r^{2 i} \qquad
(a \in \{0,1\}) ~, \label{QPol}
\end{split}\end{equation}
for some smooth integrable functions $q^{(0)}_i, q^{(1)}_i : [0,1) \to \reali$
($i \in\{0,...,N\}$).
The asymptotic expansion of $F(r)$ for large $r$ can be obtained using
the standard theory of Laplace integrals \cite{Mell3,Mell,Mell4,Mell2};
in the present subsection we only report the main results of this
analysis, making reference to Appendix \ref{AppII} for the proofs. \parn
For any $K\in\{1,2,3,...\}$ and any $v_0 \in (0,1)$, we have:
\begin{equation}\begin{split}
& F(r) = \sum_{i = 0}^{N}\!\sum_{m=0}^{K + i-1}\!
\Big[A_{i,m}(v_0,r)\!+\!B_{i,m}(v_0,r) \ln r^2\Big]r^{2(i-\aa-m-1)}
+ R_{K+1}(v_0,r) ~, \\
& \hspace{2.1cm} R_{K+1}(v_0,r) = O(r^{-2(K+\aa+1)} \ln r^2) \qquad
\mbox{for $r \to +\infty$} ~. \label{FAsySum}
\end{split}\end{equation}
In the above expression, the coefficients $A_{i,m}(v_0,r)$,
$B_{i,m}(v_0,r)$ (for $i\in\{0,...,N\}$, $m\in\{0,1,2,...\}$)
are defined by
\beq A_{i,m}(v_0,r) := q^{(0)}_{i,m}\,\ga(m\!+\!\aa\!+\!1,v_0 r^2)
+ q^{(1)}_{i,m}\,\GaL(m\!+\!\aa\!+\!1,v_0r^2) ~, \label{ABim} \feq
$$ B_{i,m}(v_0,r) := -\,q^{(1)}_{i,m}\,\ga(m\!+\!\aa\!+\!1,v_0r^2) ~, \qquad
q^{(a)}_{i,m} := \l.{d^m \over dv^m}\r|_{v=0}\!q^{(a)}_i(v) ~~(a \in\{0,1\})~, $$
where $\ga(~,~)$ denotes the lower incomplete gamma function and we put
\beq \GaL(s,z) := \int_{0}^{z}\!dw\; e^{-w}w^{s-1} \ln w \qquad\!
\mbox{for all $z\!\in\!(0,+\infty)$, $s\!\in\!\complessi$ with $\Re s > 0$} \,.
\label{DefGaL} \feq
Concerning the remainder term $R_{K+1}(v_0,r)$, we can give quantitative
estimates of the form
\beq |R_{K+1}(v_0,r)| \leqs (\Cc_{K+1} + \Dc_{K+1} \ln r^2)\,r^{-2(K+\aa+1)}
\quad \mbox{for all $r \in (0,+\infty)$} ~; \label{RCInf} \feq
these involve some constants $\Cc_{K+1}, \Dc_{K+1} > 0$,
that can be determined explicitly following the indications
of Appendix \ref{AppII}. \parn
Moreover, from the asymptotic behaviour of the incomplete gamma
$\ga(~,~)$ and of the function $\GaL(~,~)$, we infer
\begin{equation}\begin{split}
& F(r) = \sum_{i = 0}^N \sum_{m = 0}^{K+i-1} \Big[\al_{i,m}\!
+\be_{i,m}\ln r^2\Big] r^{2(i-m-\aa-1)} + \SS_{K+1}(v_0,r) ~, \\
& \hspace{1.2cm} \SS_{K+1}(v_0,r) = O(r^{-2(K+\aa+1)}\ln r^2) \quad
\mbox{for  $r\!\to\! +\infty$} ~, \label{FAsy}
\end{split}\end{equation}
where we have defined the real coefficients (for $i\!\in\!\{0,...,N\}$,
$m\!\in\!\{0,1,2,...\}$)
\begin{equation}\begin{split}
& \al_{i,m} := \Ga(m\!+\!\aa\!+\!1)\Big[q^{(0)}_{i,m}
+ \psi(m\!+\!\aa\!+\!1)\,q^{(1)}_{i,m} \Big] ~, \\
& \hspace{1.6cm} \be_{i,m} := -\Ga(m\!+\!\aa\!+\!1)\,q^{(1)}_{i,m} ~;
\label{coFAsy}
\end{split}\end{equation}
in the above definitions, $\psi(s) := \partial_s \ln\Ga(s)$ denotes
the digamma function and the $q^{(0)}_{i,m},q^{(1)}_{i,m}$ are as in
Eq. \rref{ABim}. In principle, the remainder term $\SS_{K+1}(v_0,r)$
in Eq. \rref{FAsy} could be evaluated quantitatively combining Eq. \rref{RCInf}
with some known estimates about the incomplete gamma functions; the related
computations would be very tedious, and will be avoided here. \salto
Let us now consider expression \rref{TiRenHarm} for the renormalized VEV
of the stress-energy tensor and perform inside the integrals appearing
therein the change of variable
\beq \tau = \atanh(v)~, \quad v \in (0,1) \feq
(``$\atanh$\!'' denotes the hyperbolic arctangent); in this way we obtain
$$ T^{(0)}_{\mu\nu}(r) = \int_0^{1}\!dv\;e^{- r^2\,v}\,
{(\atanh v)^{n-{d+3 \over 2}} \over 1\!-\!v^2}\,\Big[\PP^{(0)}_{\mu\nu}(\atanh v\,;r)
+ \ln(\atanh v)\,\PP^{(1)}_{\mu\nu}(\atanh v\,;r)\Big] ~, $$
\beq T^{(1)}_{\mu\nu}(r) = \int_0^{1}\!dv\;e^{- r^2\,v}\;
{(\atanh v)^{n-{d+3 \over 2}} \over 1\!-\!v^2}\;\PP^{(1)}_{\mu\nu}(\atanh v\,;r) ~.\feq
Since $\PP^{(0)}_{\mu\nu}$ and $\PP^{(1)}_{\mu\nu}$ are polynomials in
$r^2$ of degree $n\!+\!1$ (see Eq. \rref{PPPol}), we can re-express each
component of the renormalized stress-energy VEV (fixed $\mu,\nu$) in the
form \rref{FGInf}, where $Q(v;r)$ is as in Eq. \rref{QPol} with
\beq \aa := n-{d+3 \over 2} ~, \qquad N = n + 1 ~; \feq
\beq Q^{(0)}(v;r):= \feq
$$ {1 \over 1\!-\!v^2} \l({\atanh v\over v}\r)^{\!\!n-{d+3 \over 2}}
\l[\PP^{(0)}_{\mu\nu}(\atanh v\,;r) + \l(\!\ln\!\Big({\atanh v \over v}\Big)\!+ \Mk\!\r)\!
\PP^{(1)}_{\mu\nu}(\atanh v\,;r)\r] ; $$
\beq Q^{(1)}(v;r) := {1 \over 1\!-\!v^2} \l({\atanh v\over v}\r)^{\!\!n-{d+3 \over 2}}
\PP^{(1)}_{\mu\nu}(\atanh v\,;r) ~. \feq
In this way we can derive asymptotic expressions analogous to
\rref{FAsySum} (and \rref{FAsy}) for the renormalized stress-energy
VEV in the limit $r = k|\bx| \to +\infty$; we report in Section
\ref{Harmd123} the explicit results of this computation for the
cases with $d\in\{1,2,3\}$, respectively.
\vspace{-0.4cm}
\subsection{The total energy.}\label{EnSubsec} Let us recall that the
total energy can always be expressed as the sum of a bulk and a boundary
contribution as in Eq. \rref{EEtot}; in the following we are going to
discuss these two contributions separately. \parn
Let us first consider the \textsl{regularized bulk energy}; according
to Eq. \rref{defEs}, this is
$$ E^\s = {\mm^\s\! \over 2}\; \Tr\,\AA^{{1 -\s \over 2}} ~. $$
This is connected through Eq. \rref{DirHeatTr} to the heat trace $K(\t)
:= \Tr e^{-\t\AA}$ that, according to Eq. \rref{equakTr}, has the form
\beq K(\t) = {1 \over \t^d}\;H(\t) \qquad \mbox{with} \qquad
H(\t) := \l({\t \over 2\sinh(k^2\t)}\r)^{\!\!d} \,; \label{KtoHTr} \feq
it is patent that the map $\t \mapsto H(\t)$ is smooth on $[0,+\infty)$
and exponentially vanishing for $\t \to +\infty$. Then, following
the general considerations of subsection \ref{contiparts}
(see, in particular, Eq. \rref{HeatConTr} for the trace $\Tr\AA^{-s}$),
we obtain for the regularized bulk energy
\beq E^\s = {(-1)^n\,\mm^\s \over 2\,\Ga({\s - 1 \over 2})({\s - 1 \over 2}\!-\!d)
...({\s - 1 \over 2}\!-\!d\!+\!n\!-\!1)}\int_0^{+\infty} \!\!d\t \;
\t^{{\s - 3 \over 2}-d+n}\, {d^n \over d\t^n}\,H(\t) ~. \label{ERegHarm} \feq
The above relation holds for any $n \in \{1,2,3,...\}$ and the integral
appearing therein converges for any complex $\s$ with
\beq \Re\s > 2(d - n) + 1 ~; \feq
thus, for any integer $n > d + 1/2$, Eq. \rref{ERegHarm} gives the
analytic continuation of $E^\s$ in a neighborhood of $\s = 0$. Since
here no singularity appears, we can obtain the renormalized bulk
energy simply by setting $\s = 0$ in Eq. \rref{ERegHarm}; making again
the change of integration variable $\tau := k^2\t$ (see Eq. \rref{ttau}),
we infer \pagebreak
\begin{equation}\begin{split}
& E^{ren} = -{k \over 2^{d+2-n}\sqrt{\pi}}
\l(\prod_{i = 0}^{n-1}{1 \over 2(d-i)\!+\!1}\r) \int_0^{+\infty}\!\!d\tau\;
\tau^{n-d-{3 \over 2}}\,{d^n \over d\t^n}\, \HH(\tau) \\
& \hspace{1cm} \mbox{for any}\;\;n > d + {1 \over 2} ~,
\qquad \mbox{with} \quad  \HH(\tau) := \l({\tau \over \sinh \tau}\r)^{\!d} ~. \label{ERenHarm}
\end{split}\end{equation}
The above expression is fully explicit and holds for any spatial
dimention $d$; in the subsequent Section \ref{Harmd123} we are
going to evaluate numerically the integral appearing in Eq. \rref{ERenHarm}
for $d\in\{1,2,3\}$, making the minimal choice $n = d+1$\,. \parn
Let us point out that in the case under analysis, where the potential
is assumed to be isotropic, we could also derive an alternative
representation for the bulk energy $E^{ren}$ in terms of the
(analytically continued) Riemann zeta function; we discuss this
topic in detail in Appendix \ref{ExaEn}.
Here we prefer to focus the attention on the approach \rref{ERenHarm}
for the computation of $E^{ren}$ since it is more general; in fact,
it could be employed straightforwardly also when the background potential
is not isotropic
({\footnote{Let us recall that in the present paper we have chosen
to omit the analysis of cases with unisotropic background potentials
only for simplicity.}}). \parn
As pointed out in the Introduction, the computation of the renormalzied
bulk energy was also performed by Actor and Bender in \cite{ActHarm2};
these authors employed a global zeta approach based on ad hoc results
on the analytic continuation of some particular kind of ``generalized''
zeta functions (discussed in detail in \cite{ActHarm1}), befitting the
configuration under analysis. Let us stress once more that here, instead,
we derive the required analytic continuation applying a general procedure
in a mechanical way. Our method and the one of \cite{ActHarm2} could be
proved to be equivalent in any dimension, yet the general discussion is
too involved to be reported here; in subsection \ref{Harm3} we check by
direct comparison that, in the case $d = 3$
({\footnote{As a matter of fact, this is the only model discussed
both here and in \cite{ActHarm2}.}}),
the numerical result obtained using prescription \rref{ERenHarm}
agrees with the one derived in \cite{ActHarm2}.
\salto
Now, let us move on to discuss the \textsl{boundary energy}. In all the
cases treated in Parts I and II this boundary contribution was easily
found to vanish identically in consequence of the boundary conditions.
Here the situation is not so simple: since the spatial domain is unbounded
(indeed, $\Om = \reali^d$), we must resort to the procedure pointed out
in subsection \ref{TotEnSub} and intend the integral over $\partial\Om$ in
definition \rref{defBs} as the limit of integrals over the boundary of
suitable, bounded subdomains $(\Om_\ell)_{\ell = 0,1,2,...}$ (see the
comments below Eq. \rref{defBs}). Due to the spherical symmetry and to
the characteristic scale of the problem under analysis, it is natural to
choose for $\Om_\ell$ the balls with center in the origin and radius $\ell/k$:
\beq \Om_\ell := B^{(d)}_{\ell/k}(\b0) \equiv \{\bx \in \reali^d ~|~
|\bx| \leqs \ell/k \} \qquad \mbox{for $\ell \in \{0,1,2,...\}$} ~. \feq
Then, Eq. \rref{defBs} for the boundary energy reads
\beq B^\s = \mm^\s \l(\!{1 \over 4}-\xi\!\r)\!\lim_{\ell \to +\infty}
\int_{S^{(d-1)}_{\ell/k}}\!da(\bx)\,\l.\partial_{\mathfrak{r}_\by}
\Dir_{{\s + 1\over 2}}(\bx,\by)\r|_{\by = \bx} \label{BsHarm} \feq
where $S^{(d-1)}_{\ell/k} \equiv \partial B^{(d)}_{\ell/d}(\b0)$ indicates
the $(d-1)$-dimensional spherical hypersurface, while $\partial_{\mathfrak{r}_\by}$
denotes the derivative in the radial direction with respect to the variable $\by$.
In terms of the set of rescaled spherical coordinates
\rref{sphCoo}, we have
\beq \partial_{\mathfrak{r}_\by}K(\t\,;\bx,\by)\Big|_{\by = \bx}
= - \l({k \over \sqrt{2\pi \sinh(2k^2\t)}}\r)^{\!\!d}
(k\,r)\,\tanh(k^2\t)\, e^{- r^2 \tanh(k^2\t)} ~; \feq
then, using the integral representation \rref{DirHeat} for the Dirichlet
kernel in terms of the heat kernel (again, with the change of variable
$\tau := k^2 \t$) and noting that the above expression for the latter
does not depend on the angular variables $(\te_1,...,\te_{d-1})$
({\footnote{We also use the fact that the area of the
$(d-1)$-dimensional spherical hypersurface of radius $\rg$ is
$$ Area\Big(S^{(d-1)}_\rg \Big) = {2 \pi^{d/2} \over \Ga(d/2)}\; \rg^{d-1} ~. $$}}),
we infer
\beq B^\s = -\,{k \over 2^{{d \over 2}-1}\Ga({d \over 2})\,\Ga({\s + 1\over 2})}
\l({\mm \over k}\r)^{\!\!\s} \l(\!{1 \over 4}-\xi\!\r) \InB \label{BsLim} \feq
where, for brevity, we have put
\beq \InB := \lim_{\ell \to +\infty} \int_0^{+\infty}\! d\tau\;
\tau^{{\s - d + 1\over 2}} \l(e^{- \ell^2 \tanh{\tau}}\;
{\ell^d\, \tau^{d/2 - 1}\tanh{\tau} \over (\sinh(2\tau))^{d/2}}\r)
\label{defInB} \feq
For any $d \in \{1,2,3,...\}$, $\ell \geqs 0$, the expression within
the round brackets in the above equation is an analytic function of
$\tau$ on $[0,+\infty)$ and vanishes exponentially for $\tau \to + \infty$\,.
In consequence of this, we easily infer that the integral in Eq.
\rref{defInB} converges for any complex $\s$ with
\beq \Re \s > d - 3 ~. \feq
Within this region, by Lebesgue's dominated convergence theorem
({\footnote{Indeed, notice that, for any $\ell \geqs 0$, there holds
$$ \l|\tau^{{\s - d + 1\over 2}} \l(e^{- \ell^2 \tanh{\tau}}\;
{\ell^d\, \tau^{d/2 - 1}\tanh{\tau} \over (\sinh(2\tau))^{d/2}}\r)\r|
\leqs \tau^{{\s - d + 1\over 2}} \l({\tau^{d/2 - 1}\tanh{\tau} \over
(\sinh(2\tau))^{d/2}}\r) \l({d \over 2e \tanh(\tau)}\r)^{\!d/2} $$
where the right-hand side is Lebesgue-integrable over $(0,+\infty)$\,;
besides,
$$ \lim_{\ell \to +\infty} \tau^{{\s - d + 1\over 2}} \l(e^{- \ell^2 \tanh{\tau}}\;
{\ell^d\, \tau^{d/2 - 1}\tanh{\tau} \over (\sinh(2\tau))^{d/2}}\r) = 0
\qquad \mbox{for all $\tau \in (0,+\infty)$} ~. $$}})
we can take the limit under the integral sign in Eq. \rref{defInB},
and the conclusion is
\beq \InB = 0 \qquad \mbox{for $\Re \s > d- 3$} ~. \feq
Summing up, by analytic continuation we conclude that the renormalized
boundary energy vanishes:
\beq B^{ren} := B^\s\Big|_{\s = 0} = 0 ~. \feq
\section{The previous results in spatial dimension $\boma{d\in\{1,2,3\}}$}\label{Harmd123}
\subsection{Case $\boma{d=1}$.} \label{Harm1}
\begin{figure}[h!]
    \centering
        \begin{subfigure}[b]{0.49\textwidth}
                \includegraphics[width=\textwidth]{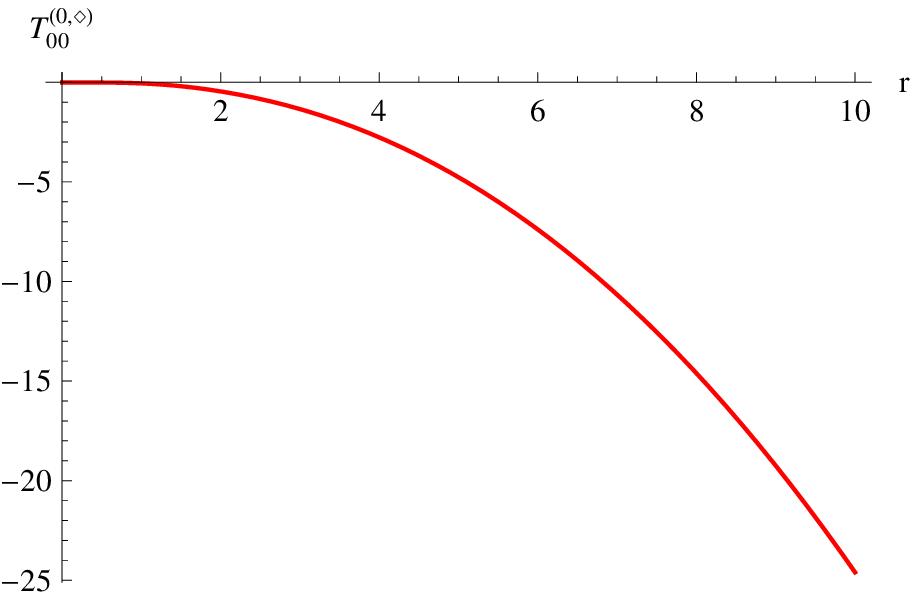}
        \end{subfigure}
        \begin{subfigure}[b]{0.49\textwidth}
                \includegraphics[width=\textwidth]{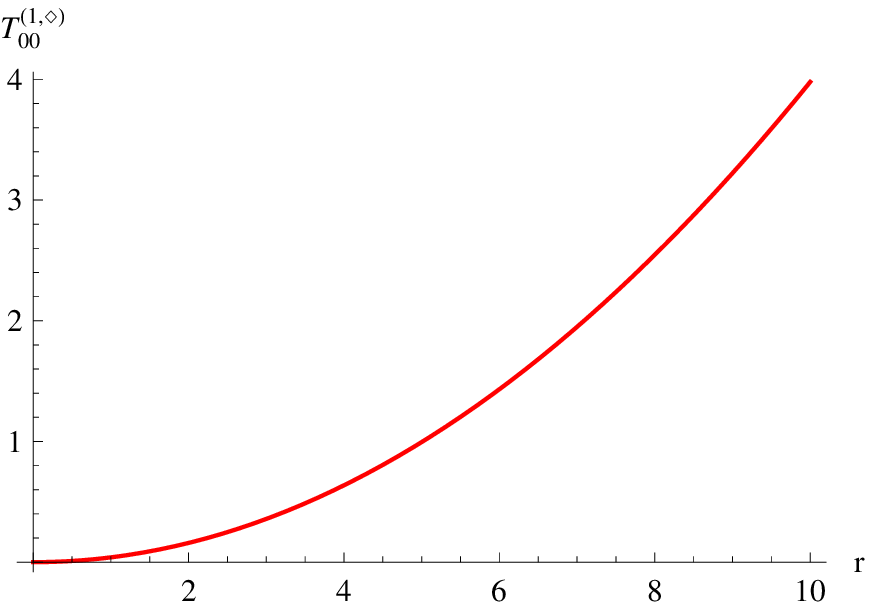}
        \end{subfigure}
        \caption{$d=1$: graphs of $T_{00}^{(0,\Co)}$ and $T_{00}^{(1,\Co)}$\,.}\label{fig:T00Co1}
\end{figure}

\begin{figure}[h!]
    \centering
        \begin{subfigure}[b]{0.49\textwidth}
                \includegraphics[width=\textwidth]{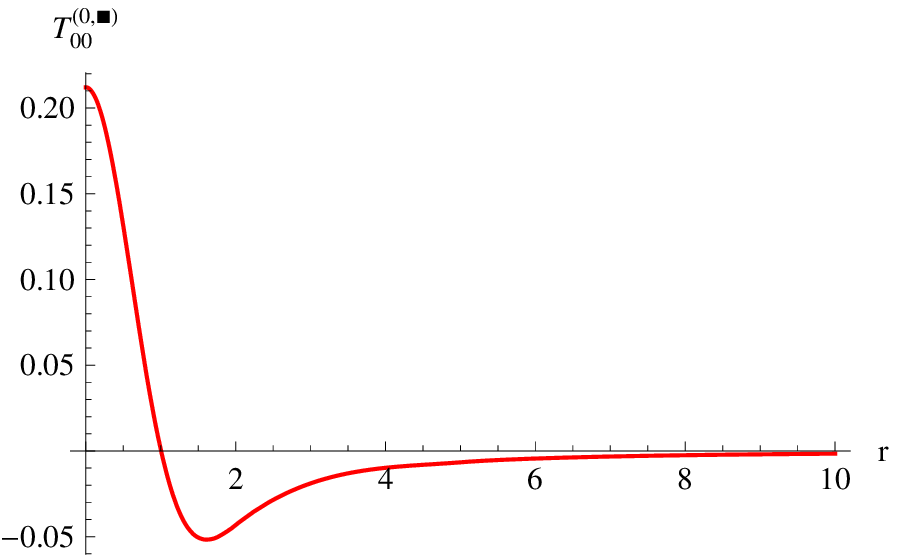}
        \end{subfigure}
        \begin{subfigure}[b]{0.49\textwidth}
                \includegraphics[width=\textwidth]{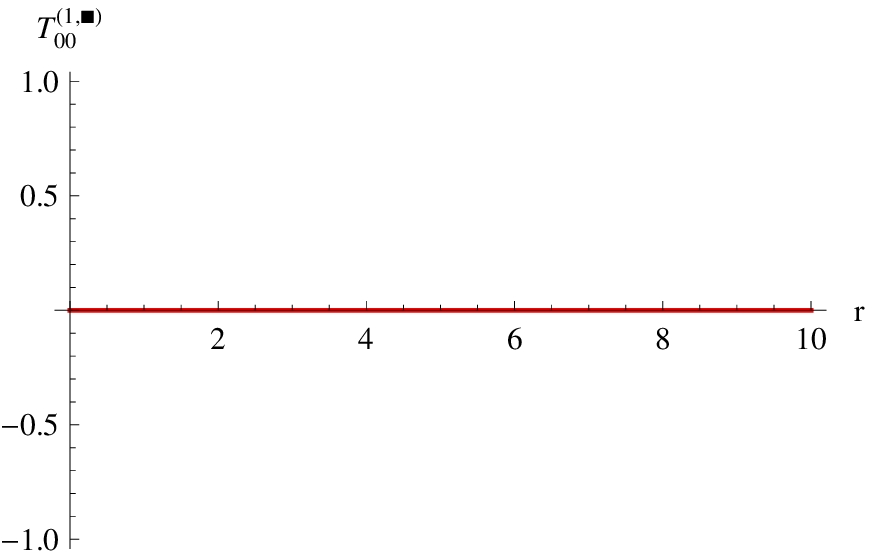}
        \end{subfigure}
        \caption{$d=1$: graphs of $T_{00}^{(0,\NCo)}$ and $T_{00}^{(1,\NCo)}$\,.}\label{fig:T00NCo1}
\end{figure}

\begin{figure}[h!]
    \centering
        \begin{subfigure}[b]{0.49\textwidth}
                \includegraphics[width=\textwidth]{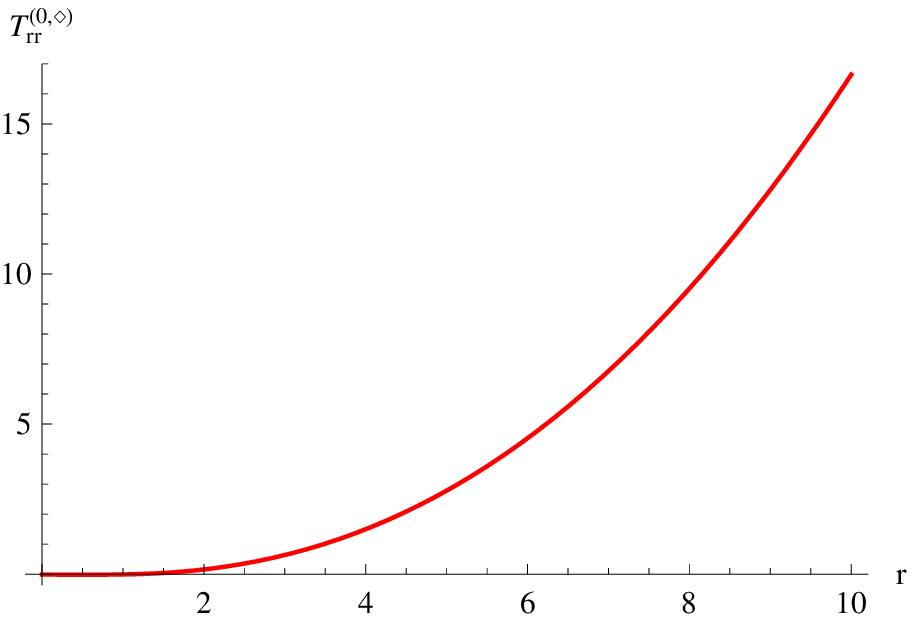}
        \end{subfigure}
        \begin{subfigure}[b]{0.49\textwidth}
                \includegraphics[width=\textwidth]{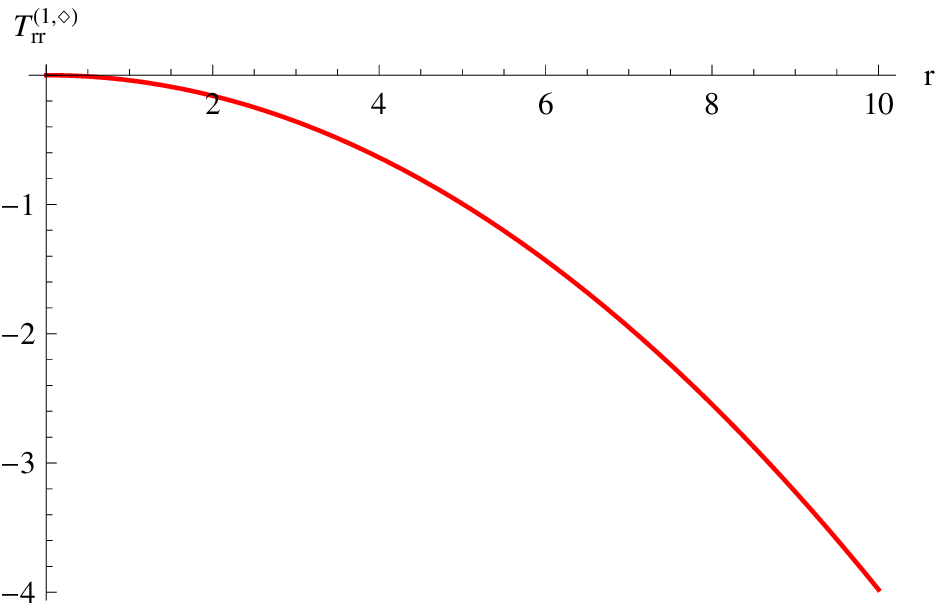}
        \end{subfigure}
        \caption{$d=1$: graphs of $T_{rr}^{(0,\Co)}$ and $T_{rr}^{(1,\Co)}$\,.}\label{fig:T11Co1}
\end{figure}

\begin{figure}[h!]
    \centering
        \begin{subfigure}[b]{0.49\textwidth}
                \includegraphics[width=\textwidth]{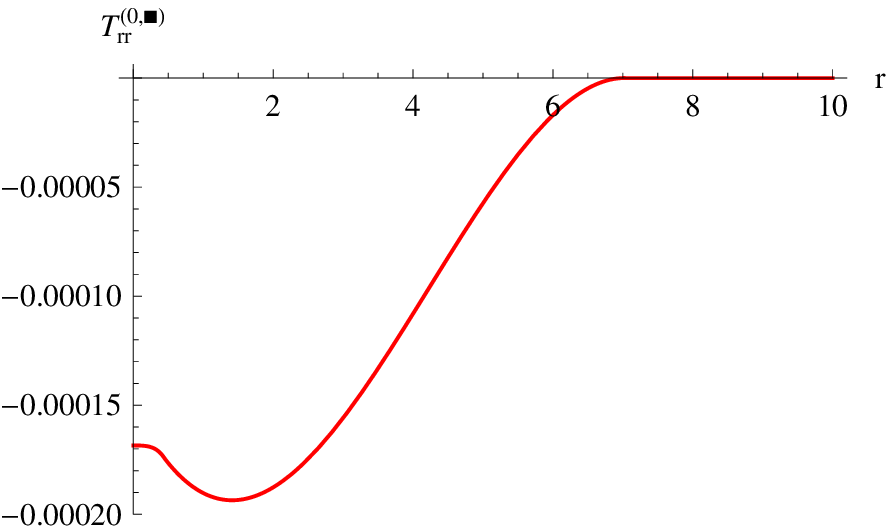}
        \end{subfigure}
        \begin{subfigure}[b]{0.49\textwidth}
                \includegraphics[width=\textwidth]{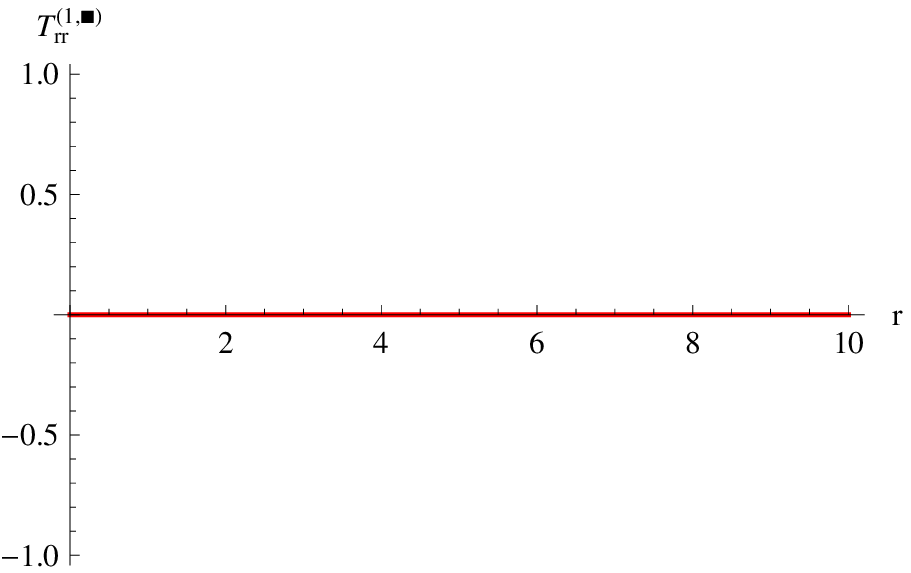}
        \end{subfigure}
        \caption{$d=1$: graphs of $T_{rr}^{(0,\NCo)}$ and $T_{rr}^{(1,\NCo)}$\,.}\label{fig:T11NCo1}
\end{figure}

Let us first recall that in this case the configuration is symmetrical
under the reflection $x^1 \to -x^1$; on any one of the intervals $(0,+\infty)$
and $(-\infty,0)$ we can use the coordinate $\bq \equiv r := k |x^1| \in
(0,+\infty)$ (see Eq. \rref{ri1}). \parn
After lenghty computations, based on Eq.s (\ref{Tidir00}-\ref{Tidirij})
and on the general considerations of subsection \ref{ACH}, we obtain
for the regularized VEV of the stress-energy tensor an expression of
the form \rref{TisHarmPol}. More precisely, we have
\beq \la 0|\Tis_{\mu\nu}(\bq)|0\ra = {k^2 \over \Ga({\s+1 \over 2})}
\l({\mm \over k}\r)^{\!\!\s}\! \int_0^{+\infty} \!\!\!d\tau\;
\tau^{-2+{\s \over 2}}\,\HH^{(\s)}_{\mu \nu}(\tau\,;\bq) \qquad
(\mu, \nu \in \{0,r\})\,, \label{TisHarm1d} \feq
where (recall that dependence on $\xi$ is understood)
\begin{equation}\begin{split}
& \hspace{5.cm} \HH^{(\s)}_{00}(\tau\,;\bq) := \\
& A_1(\tau,r)\! \l[-(1\!-\!\s)(1\!+\!4\xi) + (1\!-\!4\xi)\!
\l({2\tau \over \sinh{2\tau}}\r)\!
\l(1 + {\sinh{4\tau} \over 2\cosh^2\!\tau}\;r^2\r)\!\r]\,,
\end{split}\end{equation}
\beq \HH^{(\s)}_{rr}(\tau\,;\bq) := A_1(\tau,r)
\!\l[{2\tau\,(1\!+\!4\xi \cosh{2\tau}) \over \sinh{2\tau}}\,-(1\!-\!4\xi)\!
\l(\!1\!-\!\s+{2\tau\;r^2 \over 1\!+\!\cosh^2\!\tau}\r)\!\r]\,, \feq
and the off-diagonal components $\HH^{(\s)}_{\mu \nu}(\tau\,;\bq)$
($\mu \neq \nu$) vanish identically; in the above, for simplicity
of notation we have set
\beq A_1(\tau,r) := {1 \over 16 \sqrt{\pi}}\;
e^{-r^2 \tanh \tau}\,\sqrt{2\tau \over \sinh(2\tau)} ~. \feq
The above expressions for the components of the tensor
$\HH^{(\s)}_{\mu\nu}(\tau\,;\bq)$ are easily seen to possess the
features anticipated in Eqs. \rref{HHprel} \rref{antic} and in the
related comments. Thus, according to the general framework developed
in subsection \ref{ACH}, we can obtain the analytic continuation
of $\la 0|\Tis_{\mu\nu}(\bq)|0\ra$ given in Eq. \rref{TisHarm1d}
integrating by parts $n$ times the integral therein, with $n > 1$
(see Eq. \rref{nMin}). Let us fix
\beq n = 2 ~, \feq
so that Eq. \rref{TiHarmAC} gives
\beq \la 0|\Tis_{\mu\nu}(\bq)|0\ra = {k^2 \over \Ga({\s +1 \over 2})}
{1 \over ({\s \over 2}\!-\!1) {\s \over 2}}\l({\mm \over k}\r)^{\!\!\s}
\!\!\int_0^{+\infty}\!\!\!\!d\tau\;\tau^{\s\over 2}\,\partial_{\tau}^{2}
\HH^{(\s)}_{\mu \nu}(\tau;\bq) \,.\! \label{TiHarmAC1d} \feq
It is evident that $\la 0|\Tis_{\mu\nu}(\bq)|0\ra$ has a simple pole
at $\s = 0$ (this reflects a general feature of the cases with odd
spatial dimension, already indicated in subsection \ref{ACH}).
Thus, the renormalized stress-energy VEV is given by the regular part in
$\s = 0$ of the function in Eq. \rref{TiHarmAC1d} (see Eq. \rref{THarmRen});
with some effort, we can re-express this quantity as in Eq. \rref{TiRenHarm}, i.e.
\beq \la 0|\Ti_{\mu\nu}(\bq)|0\ra_{ren} = k^2 \Big(T^{(0)}_{\mu\nu}(r)
+ \Mk \,T^{(1)}_{\mu\nu}(r)\Big) ~, \label{TiRenHarm1d} \feq
\begin{equation*}\begin{split}
& \hspace{0.9cm} T^{(0)}_{\mu\nu}(r) := \int_0^{+\infty}\!\!\!d\tau\;
e^{-r^2\tanh\tau} \Big[\PP^{(0)}_{\mu\nu}(\tau\,;r)
+ \ln\tau\;\PP^{(1)}_{\mu\nu}(\tau\,;r)\Big] ~,\\
& T^{(1)}_{\mu\nu}(r) := \int_0^{+\infty}\!\!\!d\tau\;
e^{- r^2\tanh\tau}\;\PP^{(1)}_{\mu\nu}(\tau\,;r) ~, \qquad
\Mk := \ga_{EM}\!+\!2\ln\!\l(\!{2\mm \over k}\r) ,
\end{split}\end{equation*}
where
\begin{equation}\begin{split}
& \PP^{(0)}_{\mu \nu}(\tau\,;r) := -\,{1 \over \sqrt{\pi}}\;e^{r^2\tanh\tau}\,
\Big[\partial_\tau^2 \HH^{(0)}_{\mu \nu}(\tau;r)
+ 2\,\partial_\s \Big|_{\s = 0}\partial_\tau^2\HH^{(\s)}_{\mu \nu}(\tau;r) \Big] ~, \\
& \hspace{2cm}\PP^{(1)}_{\mu \nu}(\tau\,;r) := -\, {2 \over \sqrt{\pi}}\;
e^{r^2\tanh\tau}\;\partial_\tau^2\HH^{(0)}_{\mu \nu}(\tau;r) ~.
\end{split}\end{equation}
It is readily found that $\PP^{(0)}_{\mu \nu}$, $\PP^{(1)}_{\mu \nu}$
are polynomials of degree $N=3$ in $r^2$. We can evaluate numerically
the integrals in Eq. \rref{TiRenHarm1d}, distinguishing between the
conformal and non-conformal parts $\Co,\NCo$ of each component: see Eq. \rref{TiCNC}, recalling
that for $d = 1$ we have (see Eq. \rref{xic})
\beq \xi_1 = 0 ~. \feq
The forthcoming Fig.s \ref{fig:T00Co1}-\ref{fig:T11NCo1} show the graphs of
the functions
\beq r \mapsto T_{\mu\nu}^{(a,\times)}(r) \quad \mbox{for $\mu=\nu \in\{0,r\}$,
$a \in\{0,1\}$, $\times \in\{\Co,\NCo\}$} ~. \label{TCNC1d} \feq
\vskip -1cm \noindent
\vfill \eject \noindent
\vskip 0.1cm
\vskip 0.1cm
\vfill \eject \noindent
Let us pass to evaluate the small and large $r$ asymptotics of the
functions in Eq. \rref{TCNC1d}. On the one hand, using Eq.s (\ref{FG}-\ref{amn})
(with $N=3$) we obtain, \hbox{for $r = k |x^1| \to 0^+$,}
\begin{equation}\begin{split}
T_{00}^{(0,\Co)}(r) &= - 0.0153 + 0.0164\,r^2 - 0.0796\,r^4 + 0.0262\,r^6 +O(r^8) ~, \\
T_{00}^{(1,\Co)}(r) &= 0.0398\,r^2 +O(r^8) ~, \\
T_{00}^{(0,\NCo)}(r) &= 0.2121 - 0.3766\,r^2 + 0.2356\,r^4 - 0.0903\,r^6 +O(r^8) ~, \\
T_{00}^{(1,\NCo)}(r) &= O(r^8)~;
\end{split}\end{equation}
\begin{equation}\begin{split}
T_{rr}^{(0,\Co)}(r) &= - 0.0153 - 0.0164\,r^2 + 0.0265\,r^4 - 0.0052\,r^6 +O(r^8) ~, \\
T_{rr}^{(1,\Co)}(r) &= - 0.0398\,r^2 + O(r^8) ~, \\
T_{rr}^{(0,\NCo)}(r) &= -0.0002 - 0.0001\,r^4 + O(r^8) ~, \\
T_{rr}^{(1,\NCo)}(r) &= O(r^8) ~.
\end{split}\end{equation}
The numerical coefficients appearing here and in other small $r$ expansions
are obtained from numerical computation of the integrals in \rref{amn}. \parn
On the other hand, using Eq.s \rref{FGInf} \rref{FAsy} \rref{coFAsy}
(with $K = 3$, $\aa = 0$) we obtain the following asymptotic expansions,
for $r = k |x^1| \to +\infty$:
\begin{equation}\begin{split}
T_{00}^{(0,\Co)}(r) &= -{r^2 \over 8\pi}\Big(\ln{r^2}+\ga_{EM}+1\Big)
+ {1 \over 8\pi\,r^2} + {49 \over 120\pi\,r^6} + O(r^{-8}\ln r^2) ~, \\
T_{00}^{(1,\Co)}(r) &= {r^2 \over 8 \pi} + O(r^{-8}\ln r^2) ~, \\
T_{00}^{(0,\NCo)}(r) &= -{1 \over 2\pi\,r^2} - {5 \over 3\pi\,r^6} + O(r^{-8}\ln r^2) ~, \\
T_{00}^{(1,\NCo)}(r) &= O(r^{-8}\ln r^2) ~;
\end{split}\end{equation}
\begin{equation}\begin{split}
T_{rr}^{(0,\Co)}(r) &= {r^2 \over 8\pi}\Big(\ln{r^2}+\ga_{EM}-1\Big)
+ {1 \over 24\pi\,r^2} + {7 \over 120\pi\,r^6} + O(r^{-8}\ln r^2) ~, \\
T_{rr}^{(1,\Co)}(r) &= -{r^2 \over 8 \pi} + O(r^{-8}\ln r^2) ~, \\
T_{rr}^{(0,\NCo)}(r) &= O(r^{-8}\ln r^2) ~, \\
T_{rr}^{(1,\NCo)}(r) &= O(r^{-8}\ln r^2) ~.
\end{split}\end{equation}
Let us now discuss the bulk energy; using the expression
\rref{ERenHarm} with $n = 2$ and evaluating numerically the
integral appearing therein, we infer
\beq E^{ren} = (0.0430546469 \pm 10^{-10})\,k ~. \label{Ed1}\feq
\subsection{Case $\boma{d=2}$.} \label{Harm2}
\begin{figure}[h!]
    \centering
        \begin{subfigure}[b]{0.49\textwidth}
                \includegraphics[width=\textwidth]{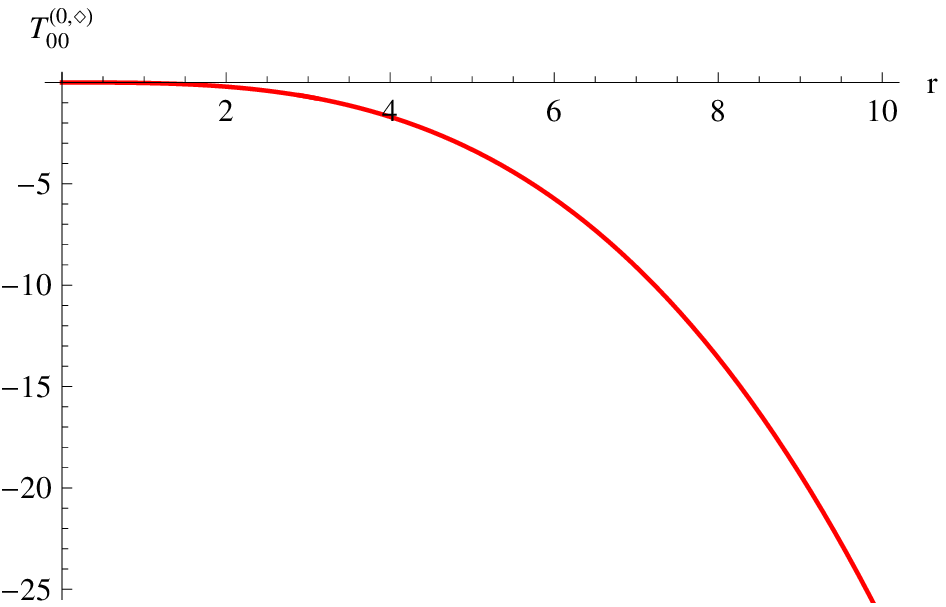}
        \end{subfigure}
        \begin{subfigure}[b]{0.49\textwidth}
                \includegraphics[width=\textwidth]{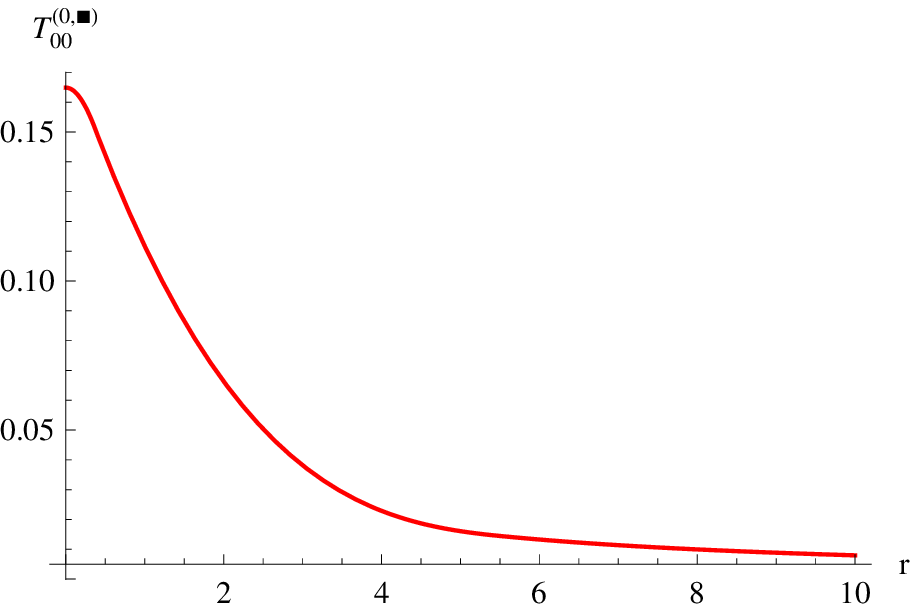}
        \end{subfigure}
        \caption{$d=2$: graphs of $T_{0 0}^{(0,\Co)}$ and $T_{0 0}^{(0,\NCo)}$\,.}\label{fig:T002}
\end{figure}

\begin{figure}[h!]
    \centering
        \begin{subfigure}[b]{0.49\textwidth}
                \includegraphics[width=\textwidth]{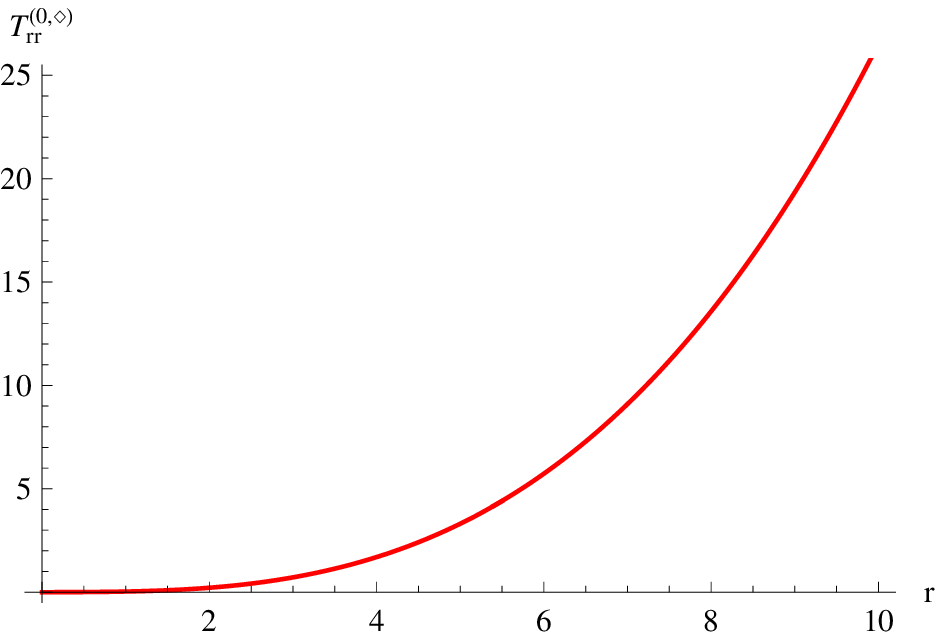}
        \end{subfigure}
        \begin{subfigure}[b]{0.49\textwidth}
                \includegraphics[width=\textwidth]{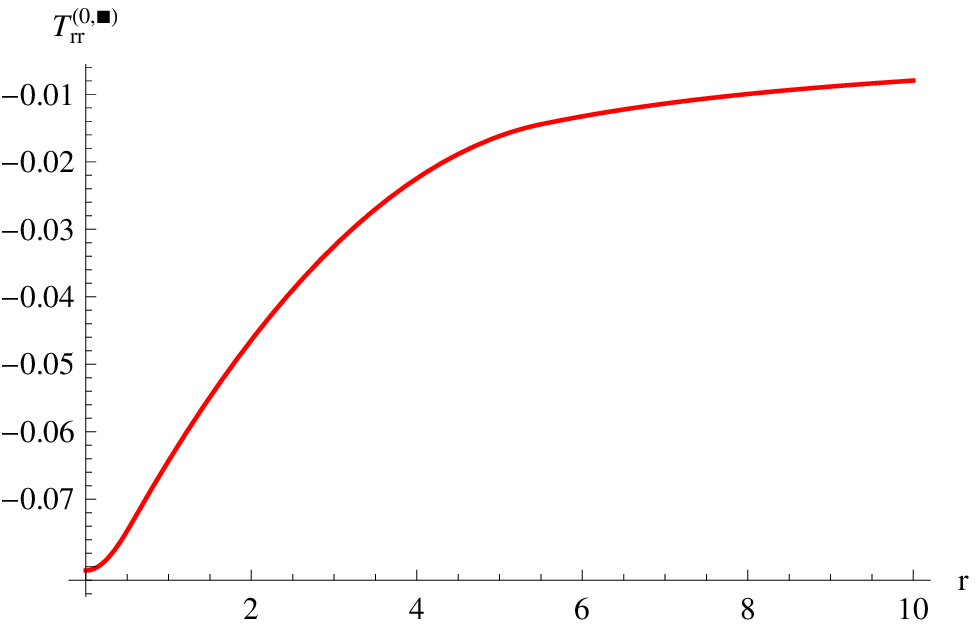}
        \end{subfigure}
        \caption{$d=2$: graphs of $T_{r r}^{(0,\Co)}$ and $T_{r r}^{(0,\NCo)}$\,.}\label{fig:T112}
\end{figure}

\begin{figure}[h!]
    \centering
        \begin{subfigure}[b]{0.49\textwidth}
                \includegraphics[width=\textwidth]{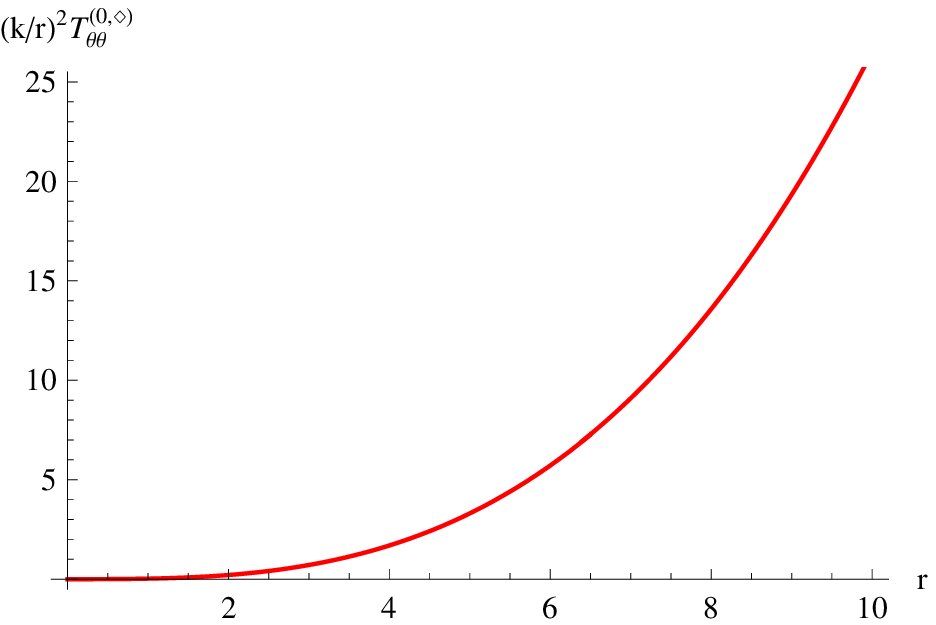}
        \end{subfigure}
        \begin{subfigure}[b]{0.49\textwidth}
                \includegraphics[width=\textwidth]{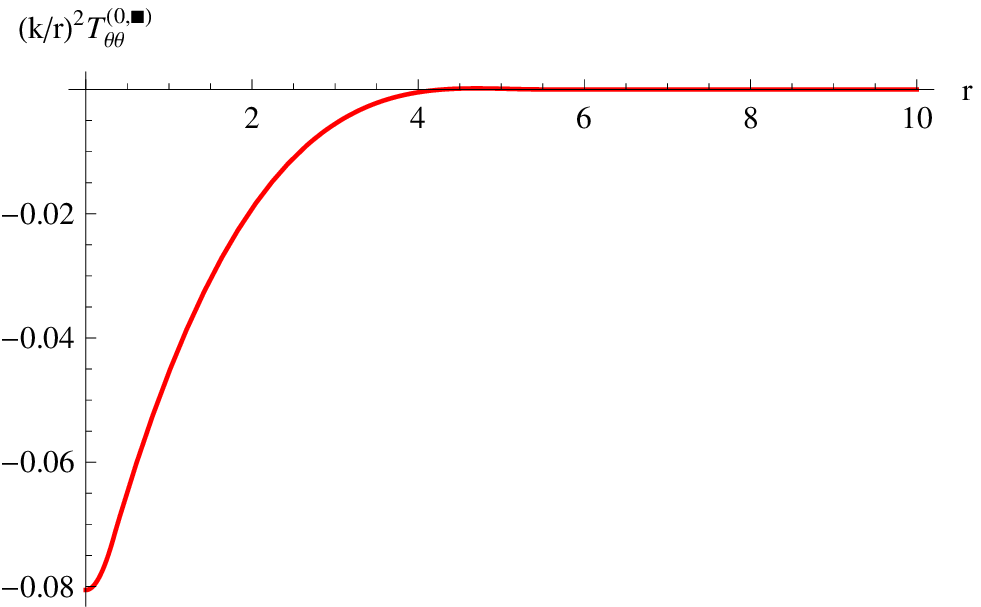}
        \end{subfigure}
        \caption{$d=2$: graphs of $(k/r)^2 T_{\te_1 \te_1}^{(0,\Co)}$
        and $(k/r)^2 T_{\te_1\te_1}^{(0,\NCo)}$\,.}\label{fig:T222}
\end{figure}

Consider the general framework of subsection \ref{ACH}; in this case
we use the (rescaled) polar coordinates $\bq = (r,\te_1)
\in (0,+\infty)\times[0,2\pi)$, fulfilling (see Eq. \rref{sphCoo})
({\footnote{Of course, the spatial line element in this coordinate system
reads $d\ell^2 = k^{-2} (dr^2\!+\!r^2 d\te^2)$; this determines the Christoffel
symbols in Eq. \rref{compu} for the derivatives $D_{ij}$.}})
\beq k\,x^1 = r \cos\te_1 ~, \qquad k\,x^2\! = r \sin\te_1~. \feq
Proceeding as in the one dimensional case, with some effort we
obtain the following integral representation for the zeta-regularized
stress-energy VEV (compare with Eq. \rref{TisHarmPol} and recall that
dependence on $\xi$ is understood):
\beq \la 0|\Tis_{\mu\nu}(\bq)|0\ra = {k^3 \over \Ga({\s+1 \over 2})}
\l({\mm \over k}\r)^{\!\!\s}\! \int_0^{+\infty} \!\!\!d\tau\;
\tau^{-{5 \over 2}+{\s\over 2}} \,\HH^{(\s)}_{\mu \nu}(\tau\,;\bq)\; \quad
(\mu,\nu \in \{0,r,\te_1\}) ~, \label{TisHarm2} \feq
where the tensor $\HH^{(\s)}_{\mu \nu}$ is diagonal
and, concerning the diagonal components, we have
\begin{equation}\begin{split}
& \hspace{5.2cm} \HH^{(\s)}_{00}(\tau\,;\bq) := \\
& A_2(\tau,r)\!\l[-(1\!-\!\s)(1\!+\!4\xi) + (1\!-\!4\xi)\!
\l({2\tau \over \sinh{2\tau}}\r)\!
\l(2+ r^2\,{\sinh{4\tau} \over 2\cosh^2\!\tau}\r)\!\r]\,, \label{HH200}
\end{split}\end{equation}
\beq \HH^{(\s)}_{rr}(\tau\,;\bq) := A_2(\tau,r)\! \l[8\xi\,{\tau \over \tanh\tau}
- (1\!-\!4\xi)\!\l(1\!-\!\s + r^2\,{2\tau \over \cosh^2\!\tau}\r)\!\r]\, , \feq
\beq \HH^{(\s)}_{\te_1 \te_1}(\tau\,;\bq) := \l({r \over k}\r)^{\!\!2} A_2(\tau,r)\!
\l[8\xi\,{\tau \over \tanh\tau} - (1\!-\!4\xi)\!\l(\!1\!-\!\s
+ r^2\,{2\tau\,\cosh{2\tau} \over \cosh^2\!\tau}\r)\!\r]\! . \feq
In the above, for simplicity of notation we have put
\beq A_2(\tau,r) := {1 \over 64\,\pi}\;
e^{-r^2 \tanh\tau} \l({2\tau \over \sinh{2\tau}}\r) \,. \label{A2} \feq
Again, the features indicated by Eq.s \rref{HHprel} \rref{antic}
(and related comments) are all possessed by the expressions (\ref{HH200}-\ref{A2})
for $\HH^{(\s)}_{\mu\nu}$. So, we can analytically continue in $\s$
the expression in Eq. \rref{TisHarm2} integrating by parts $n$ times,
with $n > 3/2$ (see Eq.s  \rref{TiHarmAC} \rref{nMin}); we choose
\beq n = 2 ~, \feq
giving
\beq \la 0|\Tis_{\mu\nu}(\bq)|0\ra = {k^3 \over \Ga({\s+1 \over 2})}\,
{1 \over ({\s\over 2}\!-\!{3 \over 2})({\s \over 2}\!-\!{1 \over 2})}
\l({\mm \over k}\r)^{\!\!\s}\!\int_0^{+\infty}\!\!\!\! d\tau\;
\tau^{{\s \over 2}-{1 \over 2}}\,\partial_{\tau}^{2}\HH^{(\s)}_{\mu \nu}(\tau;\bq) ~.
\label{TiHarmAC2d} \feq
Patently, each component of $\la 0|\Tis_{\mu\nu}|0\ra$ is regular at
$\s = 0$; thus, we can obtain the renormalized version of the
stress-energy VEV by simply putting $\s = 0$ in Eq. \rref{TiHarmAC2d}
(we already noticed this property in general for the cases with even
spatial dimension $d$; see below Eq. \rref{nMin}). In conclusion,
\beq \la 0|\Ti_{\mu\nu}(\bq)|0\ra_{ren} = k^3 \,T^{(0)}_{\mu\nu}(r)~,
\label{TiRenHarm2d} \feq
$$ T^{(0)}_{\mu\nu}(r) := \int_0^{+\infty}\!\!\!d\tau\;e^{-r^2\tanh\tau}
\,\PP^{(0)}_{\mu\nu}(\tau\,;r) ~, \quad\!
\PP^{(0)}_{\mu \nu}(\tau;r) := {4 \over 3\sqrt{\pi\,\tau}}\;
e^{r^2\tanh\tau}\; \partial_\tau^2\HH^{(0)}_{\mu \nu}(\tau;\bq) $$
(compare with Eq.s (\ref{TiRenHarm}-\ref{PP1Zer}));
also in this case, we easily check that $\PP^{(0)}_{\mu \nu}$ is a
polynomial of degree $N = 3$ in $r^2$. Next, we proceed as we did
in the previous subsection for the case of spatial dimension $d = 1$:
we evaluate numerically the integrals in Eq. \rref{TiRenHarm2d} and
separate the conformal and non-conformal parts $\Co,\NCo$
of each component (see Eq. \rref{TiCNC}), noting that Eq. \rref{xic} gives
\beq \xi_2 = {1 \over 8} ~. \label{xi2} \feq
The forthcoming Fig.s \ref{fig:T002}-\ref{fig:T222} show the graphs of
the functions
\beq r \mapsto T_{\mu\nu}^{(0,\times)}(r)\;(\mu\!=\!\nu\!\in\!\{0,r\})\,,
~ (k/r)^2\,T_{\te_1 \te_1}^{(0,\times)}(r) \qquad
\mbox{for $\times \in \{\Co,\NCo\}$} ~. \label{TCNC2d} \feq
\vfill \eject \noindent
\vfill \eject \noindent
\noindent Now, let us consider the small and large $r$ asymptotics of the
functions in \rref{TCNC2d}. On the one hand, Eq.s (\ref{FG}-\ref{amn})
(with $N=3$) give, for $r = k |\bx| \to 0^+$,
\begin{equation}\begin{split}
T_{00}^{(0,\Co)}(r) &= -0.0017 - 0.0134\,r^2 - 0.0154\,r^4 + 0.0027\,r^6 + O(r^8) ~, \\
T_{00}^{(0,\NCo)}(r) &= 0.1649 - 0.1069\,r^2 + 0.0516\,r^4 - 0.0141\,r^6 + O(r^8) ~;
\end{split}\end{equation}
\begin{equation}\begin{split}
T_{rr}^{(0,\Co)}(r) &= -0.0010 + 0.0207\,r^2 + 0.0114\,r^4 - 0.0013\,r^6 + O(r^8) ~, \\
T_{rr}^{(0,\NCo)}(r) &= -0.0806 + 0.0267\,r^2 - 0.0133\,r^4 + 0.0018\,r^6 + O(r^8) ~; \\
\end{split}\end{equation}
\begin{equation}\begin{split}
(k/r)^2\,T_{\te_1\te_1}^{(0,\Co)}(r) &= -0.0010 + 0.0140\,r^2 + 0.0153\,r^4 - 0.0027\,r^6 + O(r^8) ~, \\
(k/r)^2\,T_{\te_1\te_1}^{(0,\NCo)}(r) &= -0.0806 + 0.0802\,r^2 - 0.0440\,r^4 + 0.0124\,r^6 + O(r^8) ~. \\
\end{split}\end{equation}
Again, the above coefficients were obtained calculating numerically
the integrals in Eq. \rref{amn}. \parn
On the other hand, using Eq.s \rref{FGInf} \rref{FAsy} \rref{coFAsy}
(with $K = 5$, $\aa = -1/2$; note that no logarithmic term $\ln r^2$
appears in this case) we obtain the following
asymptotic expansions for $r = k |\bx| \to +\infty$:
\begin{equation}\begin{split}
T_{00}^{(0,\Co)}(r) &= -{r^3 \over 12 \pi} -{19 \over 2560 \pi\,r^5} + O(r^{-9}) ~, \\
T_{00}^{(0,\NCo)}(r) &= {1 \over 4\pi\,r} + {3 \over 32\pi\,r^5} + O(r^{-9}) ~;
\end{split}\end{equation}
\begin{equation}\begin{split}
T_{rr}^{(0,\Co)}(r) &= {r^3 \over 12\pi} + {1 \over 48\pi\,r} -{17 \over 2560\pi\,r^5}
+ O(r^{-9}) ~, \\
T_{rr}^{(0,\NCo)}(r) &= -{1\over 4\pi\,r} + {1 \over 32\pi\,r^5} + O(r^{-9}) ~;
\end{split}\end{equation}
\begin{equation}\begin{split}
(k/r)^2\,T_{\te_1\te_1}^{(0,\Co)}(r) &= {r^3 \over 12 \pi} - {1 \over 96\pi\,r} +
+{33 \over 2560\pi\,r^5} + O(r^{-9}) ~, \\
(k/r)^2\,T_{\te_1\te_1}^{(0,\NCo)}(r)  &= -{1 \over 8\pi\,r^5} + O(r^{-9}) ~.
\end{split}\end{equation}
As for the bulk energy, using again the expression \rref{ERenHarm}
with $n = 3$ and evaluating numerically the integral appearing therein,
we obtain
\beq E^{ren} = -(0.0180207591 \pm 10^{-10})\,k ~. \label{Ed2} \feq
\vspace{-0.9cm}
\subsection{Case $\boma{d=3}$.} \label{Harm3} 
\begin{figure}[h!]
    \centering
        \begin{subfigure}[b]{0.49\textwidth}
                \includegraphics[width=\textwidth]{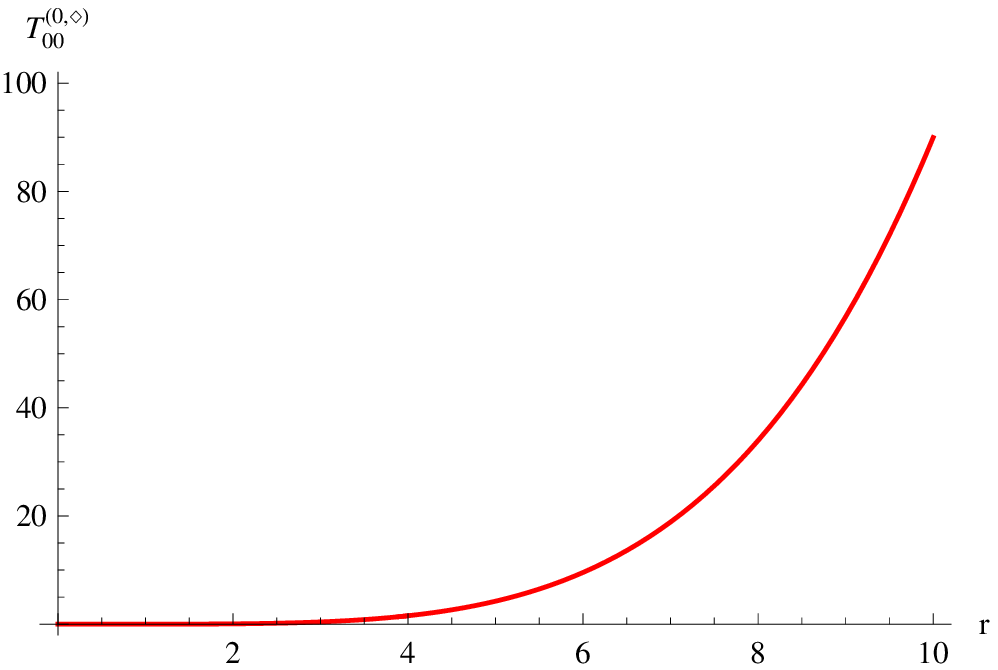}
        \end{subfigure}
        \begin{subfigure}[b]{0.49\textwidth}
                \includegraphics[width=\textwidth]{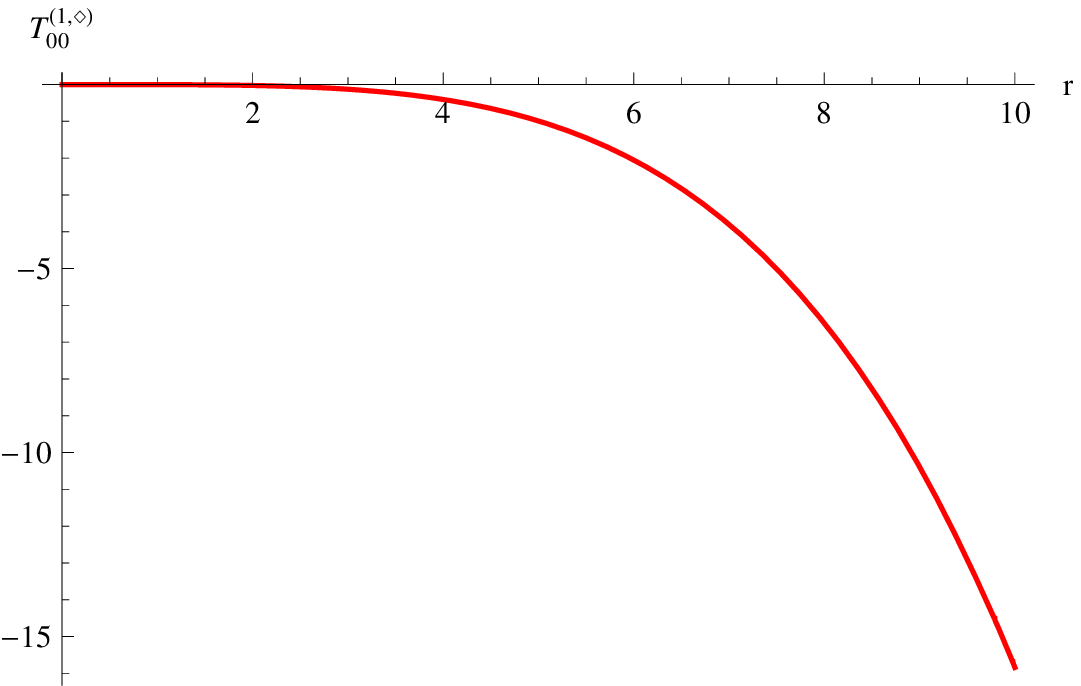}
        \end{subfigure}
        \caption{$d=3$: graphs of $T_{00}^{(0,\Co)}$ and $T_{00}^{(1,\Co)}$\,.}\label{fig:T00Co3}
\end{figure}

\begin{figure}[h!]
    \centering
        \begin{subfigure}[b]{0.49\textwidth}
                \includegraphics[width=\textwidth]{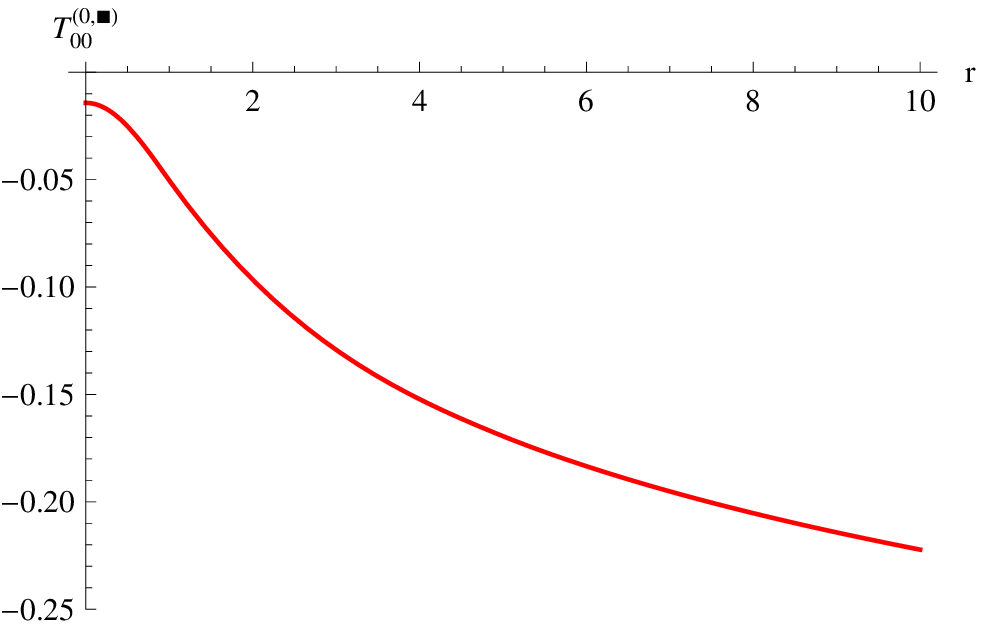}
        \end{subfigure}
        \begin{subfigure}[b]{0.49\textwidth}
                \includegraphics[width=\textwidth]{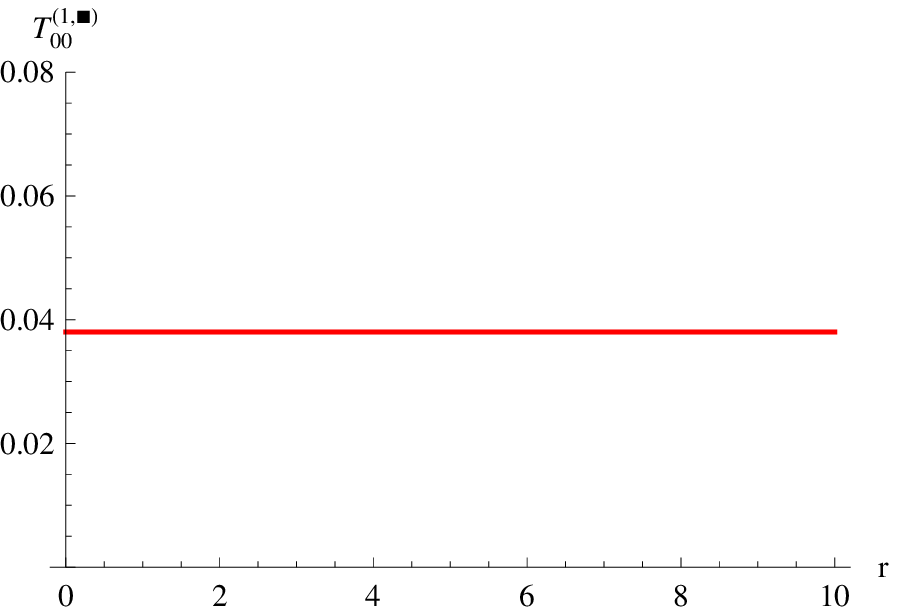}
        \end{subfigure}
        \caption{$d=3$: graphs of $T_{00}^{(0,\NCo)}$ and $T_{00}^{(1,\NCo)}$\,.}\label{fig:T00NCo3}
\end{figure}

\begin{figure}[h!]
    \centering
        \begin{subfigure}[b]{0.49\textwidth}
                \includegraphics[width=\textwidth]{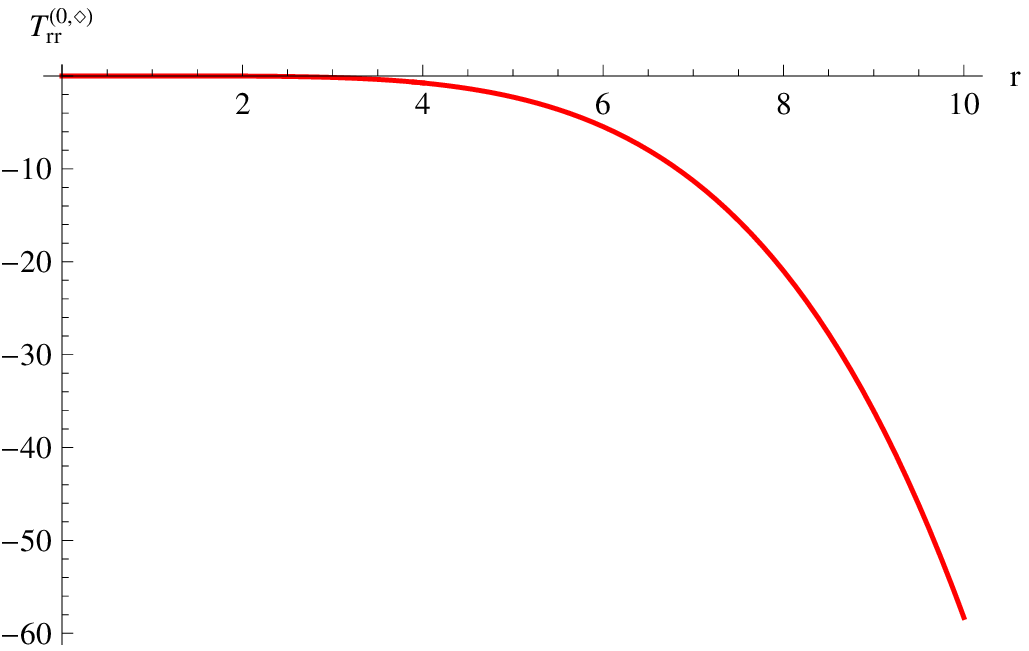}
        \end{subfigure}
        \begin{subfigure}[b]{0.49\textwidth}
                \includegraphics[width=\textwidth]{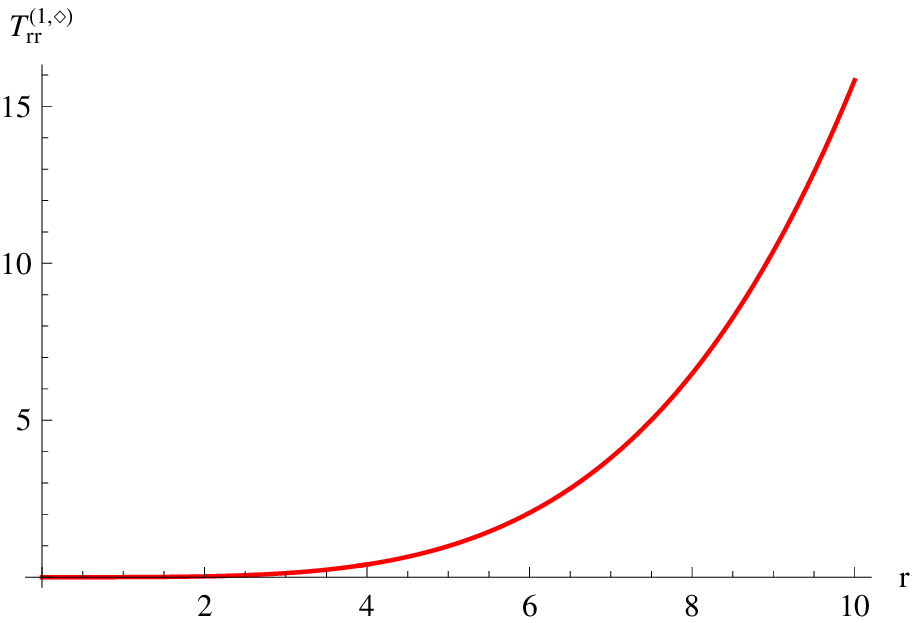}
        \end{subfigure}
        \caption{$d=3$: graphs of $T_{rr}^{(0,\Co)}$ and $T_{rr}^{(1,\Co)}$\,.}\label{fig:T11Co3}
\end{figure}

\begin{figure}[h!]
    \centering
        \begin{subfigure}[b]{0.49\textwidth}
                \includegraphics[width=\textwidth]{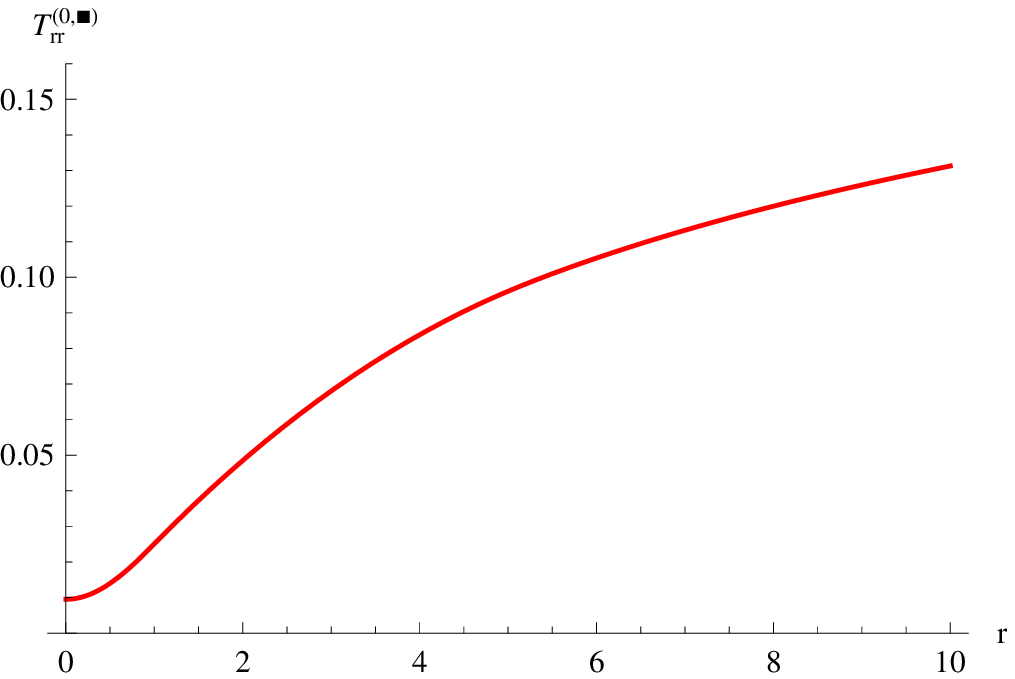}
        \end{subfigure}
        \begin{subfigure}[b]{0.49\textwidth}
                \includegraphics[width=\textwidth]{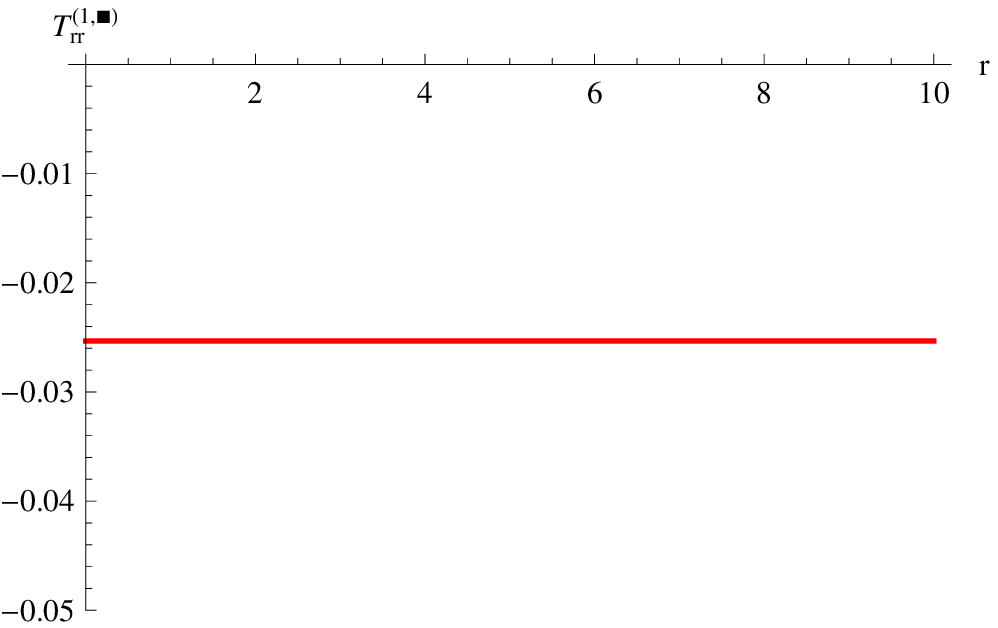}
        \end{subfigure}
        \caption{$d=3$: graphs of $T_{rr}^{(0,\NCo)}$ and $T_{rr}^{(1,\NCo)}$\,.}\label{fig:T11NCo3}
\end{figure}

\begin{figure}[h!]
    \centering
        \begin{subfigure}[b]{0.49\textwidth}
                \includegraphics[width=\textwidth]{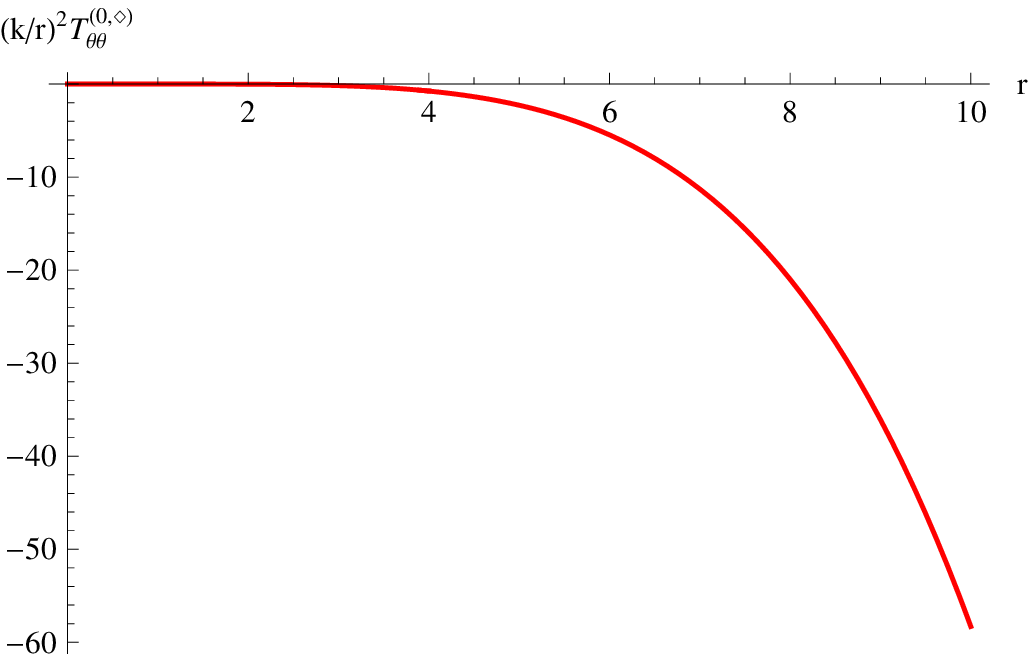}
        \end{subfigure}
        \begin{subfigure}[b]{0.49\textwidth}
                \includegraphics[width=\textwidth]{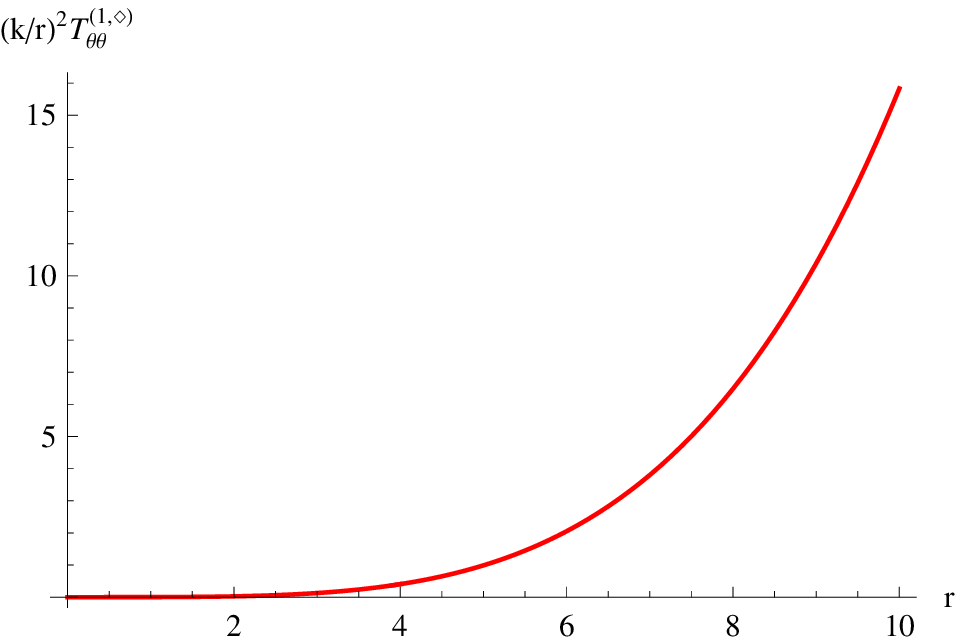}
        \end{subfigure}
        \caption{$d=3$: graphs of $(k/r)^2 T_{\te_1\te_1}^{(0,\Co)}$ and
        $(k/r)^2 T_{\te_1\te_1}^{(1,\Co)}$\,.}\label{fig:T22Co3}
\end{figure}

\begin{figure}[h!]
    \centering
        \begin{subfigure}[b]{0.49\textwidth}
                \includegraphics[width=\textwidth]{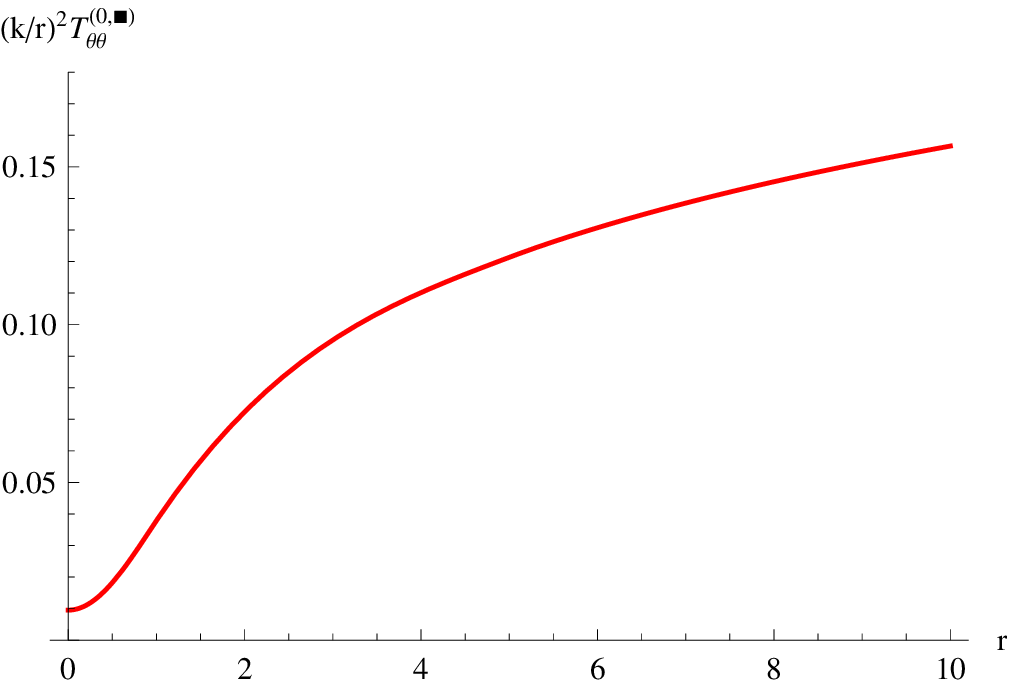}
        \end{subfigure}
        \begin{subfigure}[b]{0.49\textwidth}
                \includegraphics[width=\textwidth]{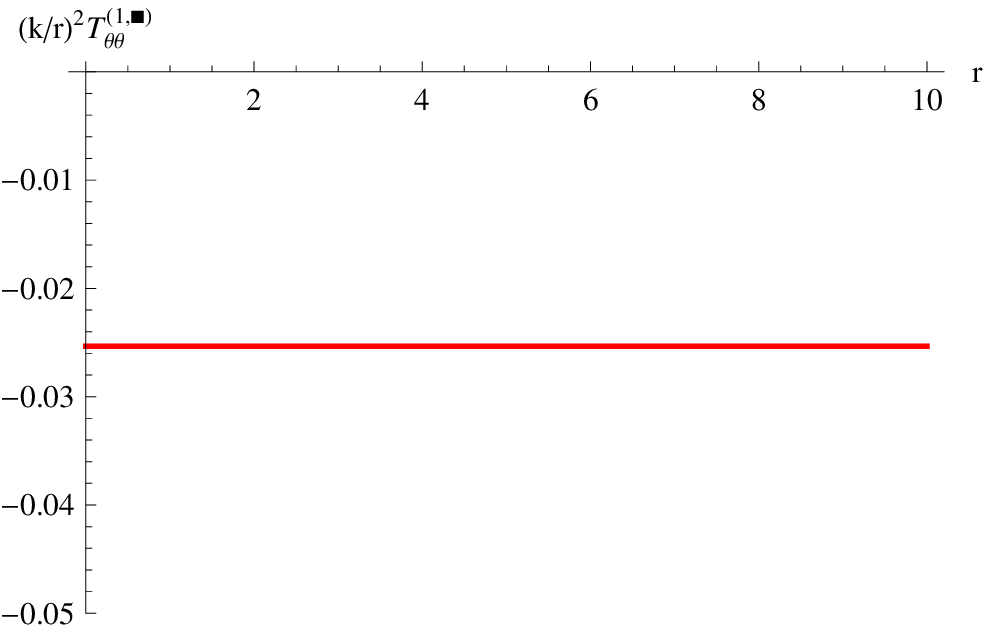}
        \end{subfigure}
        \caption{$d=3$: graphs of $(k/r)^2 T_{\te_1\te_1}^{(0,\NCo)}$
        and $(k/r)^2 T_{\te_1\te_1}^{(1,\NCo)}$\,.}\label{fig:T22NCo3}
\end{figure}

In this case we use the
coordinates $\bq = (r,\te_1,\te_2) \in (0,+\infty)\times(0,\pi)\times[0,2\pi)$,
which are related to the Cartesian coordinates $\bx \equiv (x_1,x_2,x_3)$
via (see Eq. \rref{sphCoo})
({\footnote{In this case the framework of subsection \ref{curvSubsec}
must be employed using the spatial line element $d\ell^2 = k^{-2}(dr^2\!+\!
r^2 (d\te_1^2\!+\!\sin^2\!\te_1\,d\te_2^2))$ and the corresponding
Christoffel symbols.}})
\beq k\,x^1 = r \cos\te_1 ~, \quad k\,x^2 = r \sin\te_1\cos\te_2 ~,
\quad k\,x^3 = r \sin\te_1\sin\te_2 ~. \feq
Similarly to what we did in the previous two subsections, after lenghty
computations, we can express the zeta-regularized stress-energy VEV as
(compare with Eq. \rref{TisHarmPol} and recall that dependence on $\xi$
is understood)
\beq \la 0|\Tis_{\mu\nu}(\bq)|0\ra = {k^4 \over \Ga({\s+1 \over 2})}
\l({\mm \over k}\r)^{\!\!\s}\! \int_0^{+\infty} \!\!\!\!d\tau\;
\tau^{-3+{\s \over 2}}\,\HH^{(\s)}_{\mu \nu}(\tau\,;\bq) \quad
(\mu,\nu\!\in\!\{0,r,\te_1,\te_2\})\,,\! \label{TisHarm3} \feq
where $\HH^{(\s)}_{\mu \nu}$ is diagonal and we only have to consider
the independent components
\begin{equation}\begin{split}
& \hspace{5.2cm} \HH^{(\s)}_{00}(\tau\,;\bq) := \\
& A_3(\tau,r)\!\l[-(1\!-\!\s)(1\!+\!4\xi) + (1\!-\!4\xi)\!
\l(\!{2\tau \over \sinh{2\tau}}\!\r)\!\!
\l(3+ r^2\,{\sinh{3\tau}-\sinh{\tau} \over \cosh{\tau}}\r)\r] \,, \label{HH300}
\end{split}\end{equation}
\begin{equation}\begin{split}
& \hspace{5.2cm} \HH^{(\s)}_{rr}(\tau\,;\bq) := \\
& A_3(\tau,r)\!\l[8\xi\,{\tau \over \tanh\tau} - (1\!-\!4\xi)\!
\l(1\!-\!\s + \!\l(\!{2\tau \over \sinh{2\tau}}\!\r)
\!\Big(1\!+\!2r^2 \tanh\tau\Big)\!\r)\r] \,,
\end{split}\end{equation}
\begin{equation}\begin{split}
& \hspace{5.2cm} \HH^{(\s)}_{\te_1 \te_1}(\tau\,;\bq) := \\
& \l({r \over k}\r)^{\!\!2} A_3(\tau,r)\!\l[8\xi\,{\tau \over \tanh\tau} - (1\!-\!4\xi)\!
\l(1\!-\!\s + \!\l(\!{2\tau \over \sinh{2\tau}}\!\r)
\!\Big(1\!+\!{r^2 \over \cosh^2\!{2\tau}}\Big)\!\r)\r]
\end{split}\end{equation}
(for the remaining diagonal component, i.e. $\HH^{(\s)}_{\te_2 \te_2}(\tau\,;\bq)$,
see Eq. \rref{HHprel}). In the above, for simplicity of notation we have put
\beq A_3(\tau,r) := {1 \over 64\,\pi^{3/2}}\;
e^{-r^2 \tanh\tau} \l({2\tau \over \sinh{2\tau}}\r)^{\!\!3/2}\,. \label{A3} \feq
Also this time, the expressions (\ref{HH300}-\ref{A3}) for
$\HH^{(\s)}_{\mu\nu}$ possess the properties indicated in
Eq.s \rref{HHprel} \rref{antic} (and in the related comments);
thus, we can obtain the analytic continuation in $\s$ of the
expression in Eq. \rref{TisHarm3} integrating by parts $n$ times,
for any $n > 2$ (see Eq. \rref{nMin}). For definiteness, we fix
\beq n = 3 ~, \feq
so that Eq. \rref{TiHarmAC} reads
\beq \la 0|\Tis_{\mu\nu}(\bq)|0\ra = - {k^4 \over \Ga({\s+1 \over 2})}\,
{1 \over ({\s\over 2}\!-\!2)({\s\over 2}\!-\!1){\s \over 2}}
\l({\mm \over k}\r)^{\!\!\s}\!\int_0^{+\infty}\!\! d\tau\;
\tau^{\s\over 2}\,\partial_{\tau}^{3}\HH^{(\s)}_{\mu \nu}(\tau;\bq) \,.\!
\label{TiHarmAC3d} \feq
As in all cases with odd spatial dimension, the analytic continuation of
the regularized stress-energy VEV given in Eq. \rref{TiHarmAC3d} has a
simple pole in $\s = 0$ (recall subsection \ref{ACH}).
In consequence of this, we have to adopt the extended version of the zeta
approach to define the renormalized VEV $\la 0|\Ti_{\mu\nu}(\bq)|0\ra_{ren}$,
taking the regular part in $\s = 0$ of Eq. \rref{TiHarmAC3d} (see
Eq. \rref{THarmRen}); with some effort, we obtain
\beq \la 0|\Ti_{\mu\nu}(\bq)|0\ra_{ren} = k^4 \Big(T^{(0)}_{\mu\nu}(r)
+ \Mk \,T^{(1)}_{\mu\nu}(r)\Big) ~, \label{TiRenHarm3d} \feq
\begin{equation*}\begin{split}
& \hspace{0.95cm} T^{(0)}_{\mu\nu}(r) := \int_0^{+\infty}\!\!\!d\tau\;
e^{-r^2\tanh\tau} \Big[\PP^{(0)}_{\mu\nu}(\tau\,;r)
+ \ln\tau\;\PP^{(1)}_{\mu\nu}(\tau\,;r)\Big] ~,\\
& T^{(1)}_{\mu\nu}(r) := \int_0^{+\infty}\!\!\!d\tau\;
e^{- r^2\tanh\tau}\;\PP^{(1)}_{\mu\nu}(\tau\,;r) ~, \qquad
\Mk := \ga_{EM}\!+\!2\ln\!\l(\!{2\mu \over k}\r) ,
\end{split}\end{equation*}
where
\begin{equation}\begin{split}
& \PP^{(0)}_{\mu \nu}(\tau\,;r) := -\,{1 \over 4\sqrt{\pi}}\; e^{r^2\tanh\tau}\;
\Big[3\,\partial_\tau^3 \HH^{(0)}_{\mu \nu}(\tau;\bq)
+ 4\,\partial_\s\Big|_{\s = 0}\partial_\tau^3\HH^{(\s)}_{\mu \nu}(\tau;\bq) \Big] ~, \\
& \hspace{2cm}\PP^{(1)}_{\mu \nu}(\tau\,;r) :=
-{1 \over 2\sqrt{\pi}}\;e^{r^2\tanh\tau}\;\partial_\tau^3\HH^{(0)}_{\mu \nu}(\tau;\bq) ~;
\end{split}\end{equation}
this time, we readily infer that $\PP^{(0)}_{\mu \nu}$, $\PP^{(1)}_{\mu \nu}$
are polynomials of degree $N = 4$ in $r^2$.
Now, we evaluate numerically the integrals in Eq. \rref{TiRenHarm3d} and
distinguish between the conformal and non-conformal parts
$\Co$, $\NCo$ of each component; once more we refer to Eq. \rref{TiCNC},
recalling that for $d = 3$ we have (see Eq. \rref{xic})
\beq \xi_3 = {1 \over 6} ~. \feq
The forthcoming Fig.s \ref{fig:T00Co3}-\ref{fig:T22NCo3} show the graphs of the functions
\beq r \mapsto T_{00}^{(a,\times)}(r)\,,\; T_{rr}^{(a,\times)}(r)\,,\;
(k/r)^2\,T_{\te_1 \te_1}^{(a,\times)}(r) \quad\;
\mbox{for $a\!\in\!\{0,1\}$, $\times\!\in\!\{\Co,\NCo\}$} ~. \label{TCNC3d} \feq
\vfill \eject \noindent
\vfill \eject \noindent
\vfill \eject \noindent
In conclusion, let us discuss the small and large $r$ asymptotics of the
functions in \rref{TCNC3d}. On the one hand, Eq.s (\ref{FG}-\ref{amn})
(with $N = 4$) yield, \hbox{for $r = k |\bx| \to 0^+$,}
\begin{equation}\begin{split}
T_{00}^{(0,\Co)}(r)\! &= -0.0047 - 0.0024\,r^2\! + 0.0028\,r^4\! + 0.0006\,r^6\!
- 0.0001\,r^8\! +\! O(r^{10}) \,,\! \\
T_{00}^{(1,\Co)}(r)\! &= -0.0016\,r^4\! + O(r^{10}) ~, \\
T_{00}^{(0,\NCo)}(r)\! &= -0.0143 - 0.0468\,r^2\! + 0.0134\,r^4\! - 0.0033\,r^6\!
+ 0.0007\,r^8\! +\! O(r^{10}) \,,\! \\
T_{00}^{(1,\NCo)}(r)\! &= 0.0380 + O(r^{10}) ~;
\end{split}\end{equation}
\begin{equation}\begin{split}
T_{rr}^{(0,\Co)}(r) &= -0.0016 + 0.0039\,r^2\! - 0.0003\,r^4\! - 0.0005\,r^6\! + O(r^{10}) ~, \\
T_{rr}^{(1,\Co)}(r) &= 0.0016\,r^4 + O(r^{10}) ~, \\
T_{rr}^{(0,\NCo)}(r) &= 0.0095 + 0.0188\,r^2\! - 0.0038\,r^4\! + 0.0007\,r^6\!
- 0.0001\,r^8\! + O(r^{10}) \,, \\
T_{rr}^{(1,\NCo)}(r) &= -0.0253 + O(r^{10}) ~;
\end{split}\end{equation}
\begin{equation}\begin{split}
(k/r)^2\,T_{\te_1\te_1}^{(0,\Co)}(r) &= -0.0016 + 0.0023\,r^2 + 0.0004\,r^4
- 0.0006\,r^6 + 0.0001\,r^8 + O(r^{10}) ~, \\
(k/r)^2\,T_{\te_1\te_1}^{(1,\Co)}(r) &= 0.0016\,r^4 + O(r^{10}) ~, \\
(k/r)^2\,T_{\te_1\te_1}^{(0,\NCo)}(r) &= 0.0095 + 0.0375\,r^2 - 0.0115\,r^4
+ 0.0030\,r^6 - 0.0006\,r^8 + O(r^{10}) ~, \\
(k/r)^2\,T_{\te_1\te_1}^{(1,\NCo)}(r) &= -0.0253 + O(r^{10}) ~.
\end{split}\end{equation}
$\phantom{a}$ \vspace{-0.45cm}\\
On the other hand, Eq.s \rref{FGInf} \rref{FAsy} \rref{coFAsy}
(with $K = 4$, $\aa = 0$) allow us to infer the following asymptotic
expansion, for $r = k |\bx| \to +\infty$: \vspace{-0.1cm}
\begin{equation}\begin{split}
T_{00}^{(0,\Co)}(r) &= {r^4 \over 64\pi^2}\,\Big(\ln{r^2}\!+\ga_{EM}\!+{1 \over 2}\Big)
-{5 \over 96\pi^2} -{23 \over 2880\pi^2 r^4} + O(r^{-8}\ln r^2) ~, \\
T_{00}^{(1,\Co)}(r) &= -{r^4 \over 64 \pi^2} + O(r^{-8}\ln r^2) ~, \\
T_{00}^{(0,\NCo)}(r) &= -{3 \over 8\pi^2}\,\Big(\ln{r^2\!+\!\ga_{EM}\!+{2 \over 3}}\Big)
+ {1 \over 12\pi^2 r^4} + O(r^{-8}\ln r^2) ~, \\
T_{00}^{(1,\NCo)}(r) &= {3 \over 8\pi^2} + O(r^{-8}\ln r^2) ~;
\end{split}\end{equation}
\begin{equation}\begin{split}
T_{rr}^{(0,\Co)}(r) & = -{r^4 \over 64\pi^2}\!\l(\ln{r^2}\!+\ga_{EM}\!-{3 \over 2}\r)\!
+ {1 \over 96\pi^2} - {49 \over 2880\pi^2 r^4} + O(r^{-8}\ln r^2) ~, \\
T_{rr}^{(1,\Co)}(r) &= {r^4 \over 64 \pi^2} + O(r^{-8}\ln r^2) ~, \\
T_{rr}^{(0,\NCo)}(r) &= {1 \over 4\pi^2}\,\Big(\ln{r^2}\!+\!\ga_{EM}\Big)\!
+ {1 \over 6\pi^2 r^4} + O(r^{-8}\ln r^2) ~, \\
T_{rr}^{(1,\NCo)}(r) &= -{1 \over 4\pi^2} + O(r^{-8}\ln r^2) ~;
\end{split}\end{equation}
$$ (k/r)^2\,T_{\te_1\te_1}^{(0,\Co)}(r) =  -{r^4 \over 64\pi^2}\l(\ln{r^2}\!+\ga_{EM}\!-{3 \over 2}\r)\!
- {1 \over 96\pi^2} + {31 \over 2880\pi^2 r^4} + O(r^{-8}\ln r^2) ~, $$
\begin{equation}\begin{split}
\hspace{-2.8cm} (k/r)^2\,T_{\te_1\te_1}^{(1,\Co)}(r) &= {r^4 \over 64 \pi^2} + O(r^{-8}\ln r^2) ~, \\
\hspace{-2.8cm} (k/r)^2\,T_{\te_1\te_1}^{(0,\NCo)}(r) &= {1 \over 4\pi^2}\,\Big(\ln{r^2}\!+\!\ga_{EM}\!+\!1\Big)
- {1\over 6\pi^2 r^4} + O(r^{-8}\ln r^2) ~, \\
\hspace{-2.8cm} (k/r)^2\,T_{\te_1\te_1}^{(1,\NCo)}(r) &= -{1 \over 4\pi^2} + O(r^{-8}\ln r^2) ~.
\end{split}\end{equation}
Finally, Eq. \rref{ERenHarm} with $n = 4$ and numerical evaluation
of the corresponding integral allow us to derive the bulk energy
\beq E^{ren} = -(0.0078607119 \pm 10^{-10})\,k ~. \label{EH3}\feq
This result agrees with the one obtained by Actor and Bender \cite{ActHarm2}
using a different method
({\footnote{\label{FootAct}To check this, one must compare the numerical value reported in
the above Eq. \rref{EH3} with the one reported in Eq. (4.4) of \cite{ActHarm2}.
Let us stress that conventions different from ours are used therein. In fact,
using our language, the bulk energy is formally defined in \cite{ActHarm2}
as $E := \sum_k \om_k$, while our general prescription
\rref{defEs} gives $E = {1 \over 2}\sum_k \om_k$; moreover the parameter
$\al$ of \cite{ActHarm2} and our parameter $k$ are related by $\al = \sqrt{2}\,k$.
Summing up, the ``total energy'' derived in \cite{ActHarm2} has to be multiplied
by $1/2$ in order to obtain our $E^{ren}$.}}).
\vskip 0.5cm \noindent
\textbf{Acknowledgments.}
This work was partly supported by INdAM, INFN and by MIUR, PRIN 2010
Research Project  ``Geometric and analytic theory of Hamiltonian systems in finite and infinite dimensions''.
\vfill \eject \noindent
\appendix
\section{Appendix. Asymptotic expansions for certain integrals} \label{AppII}
Hereafter we show how to obtain the expansions \rref{FserZer} \rref{FAsySum},
holding respectively for the functions defined via the integral representations
\rref{FG} \rref{FGInf}. The proofs are given in subsections \ref{Ap1} and
\ref{Ap3}; subsection \ref{Gasub} is an interlude on gamma-type functions,
useful in view of subsection \ref{Ap3}.
\vspace{-0.4cm}
\subsection{Derivation of Eq.s (\ref{FserZer}-\ref{ecn}).}\label{Ap1}
Let us consider the framework of subsection \ref{smallri}, where
\beq F(r) := \int_0^{+\infty}\!\!\!d\tau\; e^{-r^2 h(\tau)}\,P(\tau\,;r) ~,
\quad\! P(\tau\,;r) = \sum_{i = 0}^{N} p_i(\tau)\,r^{2 i}
\quad (r \in(0,+\infty)) ~, \label{FGApp} \feq
($h$ a positive bounded function, $p_i$ ($i \in\{0,...,N\}$)
some integrable functions). \parn
To evaluate $F(r)$ we start from Taylor's formula with Lagrange
remainder for the exponential; this ensures that, for any
$z \in (0,+\infty)$ and any $M \in \{0,1,2,...\}$,
\begin{equation}\begin{split}
& \hspace{1.8cm} e^{-z} = \sum_{m = 0}^M {(-z)^m\!\over m!} +
\rho_{M+1}(z_\star)\,z^{M+1} ~, \\
& \rho_{M+1}(z_\star) := {1 \over (M\!+\!1)!}\l.{d^{M+1} \over
dz^{M+1}}\,e^{-z}\r|_{z = z_\star} = {(-1)^{M+1} \over (M\!+\!1)!}\;
e^{-z_\star} ~, \label{TayExp}
\end{split}\end{equation}
for some $z_\star = z_{*}(z) \in (0,z)$; in particular, we have
\beq |\rho_{M+1}(z_\star)| \leqs {1 \over (M\!+\!1)!}~. \label{RExp}\feq
Substituting expansion \rref{TayExp} with $z = r^2 h(\tau)$ and $M = N\!-\!i$
into Eq. \rref{FGApp}, we readily infer
\begin{equation}\begin{split}
& F(r) = \sum_{i = 0}^N \l[\sum_{m = 0}^{N-i} \l({(-1)^m\!\over m!}
\int_{0}^{+\infty}\!\! d\tau\;p_i(\tau)\,(h(\tau))^m \r)\!r^{2(m + i)}\; + \r.\\
& \hspace{2.8cm}\l. + \l(\int_0^{+\infty}\!\!d\tau\;p_i(\tau)\,(h(\tau))^{N-i+1}\,
\rho_{N-i+1}(z_\star(\tau,r))\r)\!r^{2(N+1)} \r]~,
\end{split}\end{equation}
with $z_\star(\tau,r) \in (0,r^2 h(\tau))$. Introducing a new summation
index $i$ such that $m = j - i$, and performing the exchange $i \leftrightarrow j$
(\footnote{Notice as well that, for any family $(b_{ij})_{i,j = 1,...,N}$, it is
$$ \sum_{j = 0}^N \sum_{i = j}^N b_{ij} = \sum_{i = 0}^N \sum_{j = 0}^i b_{ij} ~. $$}),
the above relation yields
\beq F(r) =  \sum_{i=0}^{N} a_i \,r^{2 i} + R_{N+1}(r)
\qquad \mbox{for $r \in (0,+\infty)$} ~, \label{Fser} \feq
where the coefficients $a_i \in \reali$ are defined as in Eq. \rref{amn},
while the remainder term $R_{N+1}(r)$ is
\beq R_{N+1}(r) := \l( \sum_{i=0}^{N}\int_0^{+\infty}\!\!d\tau\;p_i(\tau)\,
(h(\tau))^{N-i+1}\, \rho_{N-i+1}(z_\star(\tau,r)) \r) r^{2(N+1)} ~. \feq
Using the estimate \rref{RExp}, we readily infer the uniform bound, holding for
all $r \in (0,+\infty)$,
\beq |R_{N+1}(r)| \leqs \left( \sum_{i=0}^{N}
{1 \over (N\!-\!i\!+\!1)!} \int_0^{+\infty}\!\!d\tau\;|p_i(\tau)|\,
(h(\tau))^{N-i+1} \right) r^{2 (N+1)}~; \label{REst} \feq
this proves Eq. \rref{RCZer}, with the expression \rref{ecn} for the constant $C_{N+1}$ therein.
\vspace{-0.4cm}
\subsection{The incomplete gamma functions and the integral $\GaL(s,x)$; asymptotics and bounds.}\label{Gasub}
Let us first consider the ``lower'' and ``upper'' incomplete gamma functions,
respectively defined as
\beq \ga(s,z) := \int_{0}^{z} dw\; e^{-w}\,w^{s-1} \qquad \mbox{for all
$s \in \complessi$ with $\Re s > 0$, $z \in (0,+\infty)$} ~, \label{LoInGa} \feq
\beq \Ga(s,z) := \int_{z}^{+\infty}\!\! dw\; e^{-w}\,w^{s-1} \qquad\mbox{for all
$s \in \complessi$, $z \in (0,+\infty)$} ~. \label{UpInGa} \feq
Hereafter we report some well-known properties of these functions (see \cite{NIST},
Chapter 8), to be used in the following subsection. First of all, there holds
\beq \ga(s,z) + \Ga(s,z) = \Ga(s) \qquad (s \notin\{0,-1,-2,...\})~, \feq
where $\Ga(~)$ is the Euler gamma function. Concerning the lower incomplete
gamma, there hold the relations
\begin{equation}\begin{split}
& \ga(s\!+\!1,z) = s\,\ga(s,z) - e^{-z} z^s ~, \quad\;
\ga(s,z) = \Ga(s)+ O(z^{-\infty})~~ \mbox{for $z \to +\infty$} ~; \\
& \hspace{3.7cm} 0 \leqs \ga(s,z) \leqs \Ga(s) \quad
\mbox{for $s \in  (0,+\infty)$} ~. \label{gaProp}
\end{split}\end{equation}
On the other hand, the upper incomplete gamma fulfills
\begin{equation}\begin{split}
& \Ga(s\!+\!1,z) = s\,\Ga(s,z) + e^{-z} z^s ~, \qquad
\Ga(s,z) = O(z^{-\infty}) ~~ \mbox{for $z \to +\infty$} ~; \\
& \hspace{3.3cm} 0 \leqs \Ga(s,z) \leqs \Ga(z) \quad
\mbox{for $s \in  (0,+\infty)$}~. \label{GaProp}
\end{split}\end{equation}
(both here and in Eq. \rref{gaProp}, the remainder $O(z^{-\infty})$ indicates
a quantity which is $O(z^{-N})$ for all $N \in \{0,1,2,...\}$). \salto
Let us move on to discuss the properties of the function $\GaL(s,x)$ defined
via the integral representation (see Eq. \rref{DefGaL})
$$ \GaL(s,z) := \int_{0}^{z}\!dw\; e^{-w}w^{s-1} \ln w \qquad\!
\mbox{for all $s \in \complessi$ with $\Re s > 0$, $z \in (0,+\infty)$} \,. $$
First of all, since $\partial_s (w^{s-1}) = w^{s-1}\ln w$, from the above
definition and Eq. \rref{LoInGa} it trivially follows that
\beq \GaL(s,z) = \partial_s\, \ga(s,z) ~; \label{derga}\feq
thus, using the recursive relation in Eq. \rref{gaProp}, we can easily infer
\beq \GaL(s\!+\!1,z) = s\,\GaL(s,z) + \ga(s,z) - e^{-z}\,z^{s}\,\ln z
\qquad (\Re s > 0) ~. \label{GalRec} \feq
Now, let us show that
\beq \GaL(1,z) = -e^{-z}\ln z -\ga_{EM} -\Ga(0,z) \label{Ga0Ex}\feq
($\ga_{EM}$ denotes the Euler-Mascheroni constant). To this purpose, let us write
\beq \GaL(1,z) = \lim_{\de \to 0^+} \GaL_{\de}(1,z) ~, \qquad \label{defGa1}
\GaL_{\de}(1,z) := \int_{\de}^{z}\! dw\; e^{-w} \ln w \quad (\de > 0) ~. \feq
Consider the regularized function $\GaL_{\de}(1,z)$ and integrate by parts to
obtain
\beq \GaL_{\de}(1,z) = -e^{-x}\ln x + e^{-\de}\ln \de
+ \int_{\de}^{z}\! dw\, e^{-w} w^{-1} ~; \feq
now, note that (for $0<\de<z$)
\beq \int_{\de}^{z}\! dw\, e^{-w} w^{-1} = \int_{\de}^{+\infty}\!\!\!dw\, e^{-w} w^{-1}
- \int_{z}^{+\infty}\!\!\! dw\, e^{-w} w^{-1} = \Ga(0,\de) - \Ga(0,z)~. \feq
Since $\Ga(0,\de) = -\ln\de  -\ga_{EM} + O(\de)$ for $\de \to 0^+$
(see \cite{NIST}, p.177, Eq. 8.4.15), we have
\beq \GaL_{\de}(1,z) = -e^{-x}\ln x -\ga_{EM} - \Ga(0,z)
- (1-e^{-\de})\ln \de + O(\de) \quad\, \mbox{for $\de \to 0^+$}, \feq
which, along with Eq. \rref{defGa1}, yields Eq. \rref{Ga0Ex}. \parn
Let us proceed to prove that there holds the bound
\beq \GaL(s,z) \geqs - s^{-2} \qquad
\mbox{for all $s,z \in (0,+\infty)$} ~; \label{GaLIneq}\feq
we are going to give the proof for $z < 1$ and $z \geqs 1$ separately. On the
one hand, for $z < 1$ (and $s>0$), there holds the following chain of inequalities:
\beq \GaL(s,z) \geqs \int_{0}^{1} dw\; e^{-w}\,w^{s-1} \ln w \geqs
\int_{0}^{1} dw\; w^{s-1} \ln w = - s^{-2} ~, \label{IneqG}\feq
where the first passage follows from the negativity of the integrand
($\ln w < 0$, for $w \in (0,1)$), while for the second we used $e^{-w} \leqs 1$
for $w \in (0,1)$. On the other hand, for $z > 1$ (and $s>0$), we have
\begin{equation}\begin{split}
\GaL(s,z) = \!\int_0^1\!dw\, e^{-w}w^{s-1} \ln w + \!\int_1^z\!dw\,e^{-w}w^{s-1} \ln w
\geqs \!\int_0^1\! dw\, e^{-w}w^{s-1} \ln w
\end{split}\end{equation}
since the integral over $(1,z)$ is positive; then, the thesis \rref{GaLIneq}
for $z > 1$ follows using the same arguments as in Eq. \rref{IneqG}. \parn
Finally, we prove the asymptotic behaviour
\beq \GaL(s,z) = \Ga(s)\,\psi(s) + O(z^{-\infty}) \qquad
\mbox{for all $s \in \complessi$ with $\Re s > 0$, $z \to +\infty$}~, \label{GaLAsy} \feq
where $s \mapsto \psi(s) := \partial_s \ln\Ga(s)$ is the digamma function.
To this purpose, write
\beq \GaL(s,z) = \int_{0}^{+\infty}\!\!\!dw\; e^{-w}w^{s-1} \ln w
- \int_{z}^{+\infty}\!\!\!dw\; e^{-w}w^{s-1} \ln w ~. \feq
Concerning the first integral, we have (see \cite{Grad}, Eq. 4.352.1)
\beq \int_{0}^{+\infty}\!\!\!dw\; e^{-w}w^{s-1} \ln w = \Ga(s)\,\psi(s)
\quad (s \in \complessi,\Re s > 0) ~, \feq
so that Eq. \rref{GaLAsy} follows if we can prove that
\beq \int_{z}^{+\infty}\!\!\!dw\; e^{-w}w^{s-1} \ln w = O(z^{-\infty})
\qquad \mbox{for $s \in \complessi$ with $\Re s > 0$, $z \to +\infty$} ~. \label{th} \feq
Indeed, for any $z \in (1,+\infty)$, we have
\beq 0 \leqs \l|\int_{z}^{+\infty}\!\!\!dw\; e^{-w}w^{s-1} \ln w\r| \leqs
\int_{z}^{+\infty}\!\!\!dw\; e^{-w}w^{\Re s} = \Ga(\Re s\!+\!1,z) ~, \feq
where the second inequality follows from the fact that $|\ln w| \leqs w$
(for $w \in (z,+\infty) \subset (1,+\infty)$), while in the last equality
we used the definition \rref{UpInGa} of the upper incomplete gamma function
(with $\Re s > 0$). Now, Eq. \rref{th} follows from Eq. \rref{GaProp}.
\vspace{-0.4cm}
\subsection{Derivation of Eq.s \rref{FAsy} \rref{coFAsy}.}\label{Ap3}
Consider the framework of subsection \ref{largeri}, where
\begin{equation}\begin{split}
& \hspace{2.cm} F(r) = \int_{0}^{1} dv\; e^{-r^2 v}\,v^{\aa}\,Q(v;r)
\qquad (\aa > -1) ~, \\
& Q(v;r) = Q^{(0)}(v;r) + Q^{(1)}(v;r) \ln v ~,\quad
Q^{(a)}(v;r) = \sum_{i = 0}^{N} q^{(a)}_i(v)\,r^{2 i} \label{FGInfApp}
\end{split}\end{equation}
for some smooth integrable functions $q^{(0)}_i, q^{(1)}_i : [0,1) \to \reali$
($i \in \{0,...,N\}$). Let us begin fixing $v_0 \in (0,1)$ and re-expressing
$F(r)$ as
\beq F(r) = \sum_{i = 0}^N \Big(I^{(<)}_{i}(v_0,r) + I^{(>)}_{i}(v_0,r)
+ J^{(<)}_{i}(v_0,r) + J^{(>)}_{i}(v_0,r)\Big)\,r^{2i} ~, \label{FIJ} \feq
where we set, for $i \in \{0,...,N\}$,
\beq I^{(<)}_i(v_0,r) := \int_{0}^{v_0}\!dv\,e^{-r^2 v}\,v^\aa q^{(0)}_i(v) ~,
\quad I^{(>)}_i(v_0,r) := \int_{v_0}^{1}\!dv\,e^{-r^2 v}\,v^\aa q^{(0)}_i(v) ~;
\label{IJ} \feq
$$ J^{(<)}_i(v_0,r) := \int_{0}^{v_0}\! dv\, e^{-r^2 v}\,v^\aa q^{(1)}_i(v) \ln v ~,
\quad J^{(>)}_i(v_0,r) := \int_{v_0}^{1}\! dv\, e^{-r^2 v}\,v^\aa q^{(1)}_i(v) \ln v ~. $$
Concerning the integrals on $(v_0,1)$, we readily infer the bounds
\begin{equation}\begin{split}
& \hspace{0.55cm} \Big|I^{(>)}_i(v_0,r)\Big| \leqs e^{-v_0\,r^2}\!
\int_{v_0}^{1}\!dv\,|v^\aa q^{(0)}_i(v)| ~, \\
& \Big|J^{(>)}_i(v_0,r)\Big| \leqs e^{-v_0 r^2}\!
\int_{v_0}^{1}\!dv\,|v^\aa \ln(v)\,q^{(1)}_i(v)| ~. \label{IJEstMa}
\end{split}\end{equation}
Let us move on to discuss the integrals $I^{(<)}_i,J^{(<)}_i$. For any
fixed $i \in\{0,...,N\}$ and any $M \in \{0,1,2,...\}$, consider the Taylor
expansions near $v = 0$ with Lagrange remainder
\beq  q^{(a)}_i(v) = \sum_{m = 0}^{M} q^{(a)}_{i,m}\,v^m
+ \rho^{(a)}_{i,M+1}(v_{\star})\,v^{M+1} ~; \label{TayR} \feq
$$ \rho^{(a)}_{i,M+1}(v_{\star}) := {1 \over (M\!+\!1)!}\l.{d^{M+1} \over
dz^{M+1}}\,q^{(a)}_i(v)\r|_{v = v_\star} ~, \qquad \mbox{for some
$v_\star = v_\star(v) \in (0,v)$} ~. $$
Of course,
\beq {~}\hspace{-0.4cm} \sup_{v \in (0,v_0)}\!\!|\rho^{(a)}_{i,M+1}(v_\star)|
\leqs {s^{(a)}_{i,M+1}(v_0) \over (M\!+\!1)!}\,, \quad
\mbox{with} ~ s^{(a)}_{i,M+1}(v_0) := \! \sup_{v \in (0,v_0)}\!
\l|{d^{M+1} \over dv^{M+1}}\, q^{(a)}_i(v)\r|. \label{estRes} \feq
Now, substitute the expansion \rref{TayR} into Eq. \rref{IJ} and make
the change of variable $v = w/r^2$ (with $w\!\in\!(0,v_0\,r^2)$) in
the $(<)$ integrals appearing therein. Recalling the definitions \rref{LoInGa}
and \rref{DefGaL} respectively of the lower incomplete gamma $\ga(~,~)$
and of the function $\GaL(~,~)$, we obtain
({\footnote{For example, we have the following chain of equalities:
$$ \int_{0}^{v_0}\!\! dv\;e^{-r^2 v} v^{m+\aa} \ln v = r^{-2(m+\aa+1)}\!
\int_{0}^{v_0 r^2}\!\!\! dw\;e^{-w} w^{m+\aa} \ln (w/r^2) = $$
$$ = r^{-2(m+\al+1)}\!\!\!\int_{0}^{v_0 r^2}\!\!\!\!\!dw\, e^{-w} w^{m+\aa}
\Big[\ln w - \ln r^2\Big]\! = r^{-2(m+\aa+1)}\Big[\GaL(m\!+\!\aa\!+\!1,v_0\,r^2)
-\ga(m\!+\!\aa\!+\!1,v_0 r^2)\ln r^2\Big] .$$ }})
\beq I^{(<)}_i(v_0,r) = \sum_{m = 0}^{M} q^{(0)}_{i,m} \,
\ga(m\!+\!\aa\!+\!1, v_0\,r^2)\;r^{-2(m+\aa+1)} + R^{(0)}_{i,M+1}(v_0,r) ~,
\label{IAs} \feq
\beq J^{(<)}_i(v_0,r) = \label{JAs} \feq
$$ = \sum_{m = 0}^{M} q^{(1)}_{i,m}
\Big[\GaL(m\!+\!\aa\!+\!1,v_0\,r^2) - \ga(m\!+\!\aa\!+\!1, v_0\,r^2)\ln r^2\Big]
r^{-2(m+\aa+1)} + R^{(1)}_{i,M+1}(v_0,r) ~, $$
where the remainder functions are defined as
\begin{equation}\begin{split}
& \hspace{0.4cm} R^{(0)}_{i,M+1}(v_0,r) := \int_{0}^{v_0}\!dv\;
e^{-r^2 v}\;\rho^{(0)}_{i,M+1}(v_{\star})\;v^{M+\aa+1} ~, \\
& R^{(1)}_{i,M+1}(v_0,r) := \int_{0}^{v_0}\!dv\;e^{-r^2 v}\;
\rho^{(1)}_{i,M+1}(v_{\star})\;v^{M+\aa+1} \ln v ~.
\end{split}\end{equation}
Then, using the bound \rref{estRes} and the definitions \rref{DefGaL}
\rref{LoInGa}, we readily infer the estimates
\beq |R^{(0)}_{i,M+1}(v_0,r)| \leqs {s^{(0)}_{i,M+1}(v_0)\over (M\!+\!1)!}\;
\ga(M\!+\!\aa\!+\!2, v_0\,r^2)\,r^{-2(M+\aa+2)} ~, \feq
$$ |R^{(1)}_{i,M+1}(v_0,r)| \leqs {s^{(1)}_{i,M+1}(v_0)\over (M\!+\!1)!}\,
\Big[\ga(M\!+\!\aa\!+\!2, v_0\,r^2)\ln r^2 - \GaL(M\!+\!\aa\!+\!2,v_0\,r^2)\Big]
r^{-2(M+\aa+2)} \,; $$
these, along with the bounds in Eq.s \rref{gaProp} \rref{GaLIneq}, imply in turn
\beq |R^{(0)}_{i,M+1}(v_0,r)| \leqs \l({\Ga(M\!+\!\aa\!+\!2)\over
(M\!+\!1)!} \,s^{(0)}_{i,M+1}(v_0)\!\r) r^{-2(M+\aa+2)} ~, \label{ResI}\feq
$$ |R^{(1)}_{i,M+1}(v_0,r)| \leqs \l({\Ga(M\!+\!\aa\!+\!2)\over (M\!+\!1)!}\,
s^{(1)}_{i,M+1}(v_0)\!\r)\!\! \l[\ln r^2\! + {(M\!+\!1)! \over
(M\!+\!\aa\!+\!2)^2}\r]r^{-2(M+\aa+2)} ~. $$
Summing up, Eq.s \rref{FAsySum} \rref{ABim} follow from Eq. \rref{FIJ} and
the relations in Eq.s \rref{IAs} \rref{JAs} with $M = K+i-1$; the remainder
estimate \rref{RCInf} descend easily from Eq.s \rref{IJEstMa} \rref{ResI},
which also give explicit expressions for the constants in \rref{RCInf}.
Finally, using the asymptotic expansions in Eq.s \rref{gaProp} \rref{GaLAsy}
(respectively holding for the lower incomplete gamma and for the function
$\GaL(~,~)$), Eq.s \rref{IAs} \rref{ResI} imply Eq.s \rref{FAsy} \rref{coFAsy}.
\section{Appendix. An alternative representation for the bulk energy} \label{ExaEn}
In subsection \ref{EnSubsec} we obtained an integral representation for
the renormalized bulk energy $E^{ren}$ (see Eq. \rref{ERenHarm}) to be
evaluated numerically
({\footnote{We pointed out in the cited subsection that the main advantage
of this approach is that it work also for configurations where the
background harmonic potential is not isotropic.}}).
In the present appendix we derive an alternative representation for the
regularized energy $E^{\s}$, allowing to express it in terms of the Riemann
zeta function (see \cite{NIST}, p. 602, Eq. 25.2.1)
\beq \zeta(s) := \sum_{n = 1}^{+\infty} {1 \over n^s} \qquad
\mbox{for $s \in \complessi$ with $\Re s > 1$} ~. \feq
Next, the renormalized energy $E^{ren}$ is computed via the zeta approach,
using the well-known analytic continuation of $\zeta$ to the whole complex
plane. \parn
Let us stress that the methods discussed here only works if the background
potential is isotropic. Nonetheless, they can be of some interest since,
for example, they allow us to perform a direct comparison with the results
of Actor and Bender \cite{ActHarm2} in the case with $d = 3$ (see subsection
\ref{IAp3}). \parn
In the forthcoming subsection \ref{IAp1} we introduce a family of integrals
and discuss some relations allowing to express them in terms of the Riemann
zeta; in the following subsection \ref{IAp2} the regularized bulk energy is
represented in terms of these integrals and its analytic continuation is
discussed. In the concluding subsection \ref{IAp3} we use the results of
subsections \ref{IAp1} \ref{IAp2} to derive, as examples, the renormalized
energy $E^{ren}$ in the cases with $d \in \{1,2,3\}$\,.
\vspace{-0.4cm}
\subsection{The integrals $\boma{\IE_n(s)}$ and a recursive relation.}\label{IAp1}
Let us consider the family of integrals
\beq \IE_n(s) := \int_0^{+\infty}\!d\tau\;{\tau^{s-1} \over \sinh^n\!\tau} \qquad
\mbox{for $n\!\in\!\{1,2,3,...\}$, $s\!\in\!\complessi$ with $\Re s > n$}
\label{IEdef} \feq
(of course, the restriction on $\Re s$ arises in order to guarantee the
convergence of the integral $\IE_n$). \parn
First note that, for $n = 1$ and $n = 2$, the above integrals are known to
be related to the Riemann zeta function $\zeta$; more precisely, we have
(see \cite{NIST}, p.604, Eq.s 25.5.8 and 25.5.9)
\beq \IE_1(s) = 2(1\!-\!2^{-s})\,\Ga(s)\,\zeta(s) ~, \label{IEEx1} \feq
\beq \IE_2(s) = 2^{-(s-2)}\,\Ga(s)\,\zeta(s-1) ~. \label{IEEx2} \feq
Next, let us fix $n \in \{1,2,3,...\}$ and consider the elementary identity
\beq {d^2 \over d\tau^2}\!\l({1 \over \sinh^n\!\tau}\r)\! =
{n(n\!+\!1) \over \sinh^{n+2}\!\tau} + {n^2 \over \sinh^n\!\tau} ~; \feq
using this result and the definition \rref{IEdef}, we infer
(for $s \in \complessi$ with $\Re s > n+2$)
\beq \IE_{n+2}(s) = -\,{n \over n\!+\!1}\;\IE_n(s)
+ {1 \over n(n\!+\!1)}\int_0^{+\infty}\!d\tau\;
\tau^{s-1}\,{d^2 \over d\tau^2}\!\l({1 \over \sinh^n\!\tau}\r) \,.
\label{IE2} \feq
Concerning the integral appearing in the right-hand side of the above
equation, integrating by parts two times we obtain
\begin{equation}\begin{split}
& \hspace{2.2cm} \int_0^{+\infty}\!d\tau\;\tau^{s-1}\,
{d^2 \over d\tau^2}\!\l({1 \over \sinh^n\!\tau}\r) = \\
& (s-1)(s-2)\!\int_0^{+\infty}\!d\tau\;{\tau^{s-3} \over \sinh^n\!\tau} =
(s-1)(s-2)\,\IE_{n}(s-2) ~. \label{IntIE2}
\end{split}\end{equation}
Summing up, Eq.s \rref{IE2} \rref{IntIE2} give the recursive relation
\beq \IE_{n+2}(s) = -\,{n \over n\!+\!1}\;\IE_n(s)
+ {(s-1)(s-2) \over n(n\!+\!1)}\;\IE_{n}(s-2) \quad
(s \in \complessi,\,\Re s > n + 2) ~. \label{RecIE} \feq
This result and the explicit expressions \rref{IEEx1} \rref{IEEx2}
allow to express any integral $\IE_n(s)$ in terms of (suitable linear
combinations of) the Riemann zeta $\zeta$; needless to say,
the analytic continuation of $\IE_n$ is then determined by the well-known
analytic continuation of $\zeta$.
\vspace{-0.4cm}
\subsection{The bulk energy in terms of the integrals $\boma{\IE_n(s)}$.} \label{IAp2}
Using Eq. \rref{ERegHarm} for the regularized energy $E^{\s}$ with $n = 0$
and the explicit representation \rref{KtoHTr} for the heat trace $K(\t)$,
we obtain
\beq E^\s = {\mm^\s \over 2^{d+1}\,\Ga({\s - 1 \over 2})}\int_0^{+\infty}\!\!d\t\;
{\t^{\s - 3 \over 2} \over \sinh^d(k^2\t)} ~; \feq
making the change of variable $\tau := k^2 \t \in (0,+\infty)$ and
recalling the definition \rref{IEdef}, we infer
\beq E^\s = {k \over 2^{d+1}\,\Ga({\s - 1 \over 2})}\l({\mm \over k}\r)^{\!\!\s}\,
\IE_{d}\!\l({\s - 1 \over 2}\r) \,. \label{IEu} \feq
The recursive relation \rref{RecIE} for $\IE_d$ and the identities
\rref{IEEx1} \rref{IEEx2} allow us to determine the analytic continuation
of $E^{\s}$ to the whole complex plane via Eq. \rref{IEu}. \parn
The renormalized bulk energy $E^{ren}$ is defined according to Eq. \rref{ERegnGen}, setting
$$ E^{ren} := RP \Big|_{\s=0} E^\s \qquad \l(\mbox{or } \;
E^{ren} := E^\s \Big|_{\s=0}\,\r)\,. $$
The final expression of $E^{ren}$ for arbitrary spatial dimension $d$
is too complicate to be reported here; in the next subsection we
consider, as examples, the cases with $d \in \{1,2,3\}$\,.
\vspace{-0.4cm}
\subsection{The bulk energy in terms of the Riemann zeta function for $\boma{d\!\in\!\{1,2,3\}}$.} \label{IAp3}
As mentioned at the end of the previous subsection, here we exemplify
the general methods developed in the previous subsections \ref{IAp1} \ref{IAp2}
for the computation of $E^{ren}$ in the cases with $d \in \{1,2,3\}$;
we also show that the results found are in agreement with those derived
in Section \ref{Harmd123} using another approach. \parn
\textsl{The case $d = 1$}. Using Eq.s \rref{IEEx1} \rref{IEu} and setting
$\s = 0$, we infer
\beq E^{ren} = -k\l({\sqrt{2}\!-\!1 \over 2}\r)\zeta\!\l(\!-{1 \over 2}\r)\,.
\label{IEu1} \feq
Evaluating numerically this expression, we have (in agreement with Eq. \rref{Ed1})
\beq E^{ren} = k(0.0430546468...) ~. \label{IEu1N} \feq
\textsl{The case $d = 2$}. Eq.s \rref{IEEx2} and \rref{IEu} with $\s = 0$ imply
\beq E^{ren} = {k \over \sqrt{2}}\;\zeta\!\l(\!-{3 \over 2}\r)\,. \label{IEu2} \feq
Numerical evaluation of the above result gives
\beq E^{ren} = -k(0.0180207590...) ~, \label{IEu2N} \feq
which agrees with the one of Eq. \rref{Ed2} (the last digit being different
due to truncation approximation). \parn
\textsl{The case $d = 3$}. This time we have to resort to the recursive
relation \rref{RecIE} (here employed with $n = 1$) as well as to the
identity \rref{IEEx1}. Then, using once more Eq. \rref{IEu} with $\s = 0$,
we infer
\beq E^{ren} = k \l[\l({\sqrt{2}\!-\!1 \over 16}\r)\zeta\!\l(\!-{1 \over 2}\r)
- \l({4\sqrt{2}\!-\!1 \over 16}\r)\zeta\!\l(\!-{5 \over 2}\r)\r]\,. \label{IEu3} \feq
We state that the above result coincides with the one of Eq. (4.4) in \cite{ActHarm2},
which in our language reads (see footnote \ref{FootAct})
\beq E^{ren} = k \l[{1 \over 2 \sqrt{2}}\,\zeta\!\l(\!-{5 \over 2}\,,{3\over 2}\r)
- {1 \over 8 \sqrt{2}}\,\zeta\!\l(\!-{1 \over 2}\,,{3 \over 2}\r)\r] \feq
where $\zeta(~,~)$ indicates the analytic continuation of the Hurwitz zeta
function; this statement follows is easily proved using the known identities
(see \cite{NIST}, p. 607, Eq. 25.11.3 and p. 608, Eq. 25.11.11)
\beq \zeta(s,a+1) = \zeta(s,a) - a^{-s} ~, \feq
\beq \zeta\!\l(s,{1 \over 2}\r) = (2^s-1)\,\zeta(s) ~. \feq
Finally, let us note that evaluating numerically the expression in the
right-hand side of Eq. \rref{IEu3} we have
\beq E^{ren} = - k (0.0078607118...) ~, \label{IEu3N} \feq
in agreement with Eq. \rref{EH3}.

\vfill \eject \noindent


\end{document}